\pdfoutput=1

\documentclass[11pt]{article}

\usepackage[preprint]{acl}

\usepackage[T1]{fontenc}
\usepackage{inconsolata}

\usepackage{times}
\usepackage{latexsym}

\usepackage[utf8]{inputenc}

\usepackage{microtype}

\usepackage{inconsolata}

\usepackage{graphicx}

\usepackage{booktabs}
\usepackage{amsmath}
\usepackage{amssymb}
\usepackage{comment}
\usepackage{multirow} 

\usepackage{subcaption}

\usepackage{xspace}
\newcommand{\ccnl}{\texttt{CodeChange2NL}\xspace}
\usepackage{xcolor}

\usepackage[most]{tcolorbox}

\definecolor{backcolour}{rgb}{0.95,0.95,0.95}
\definecolor{codegreen}{rgb}{0, 0.6, 0}          
\definecolor{codegray}{rgb}{0.5, 0.5, 0.5}       
\definecolor{codepurple}{rgb}{0.58, 0, 0.82}     

\usepackage{listings}
\lstdefinestyle{mystyle}{
  aboveskip=3mm,
  showstringspaces=false,
  columns=flexible,
  numbers=none,
  backgroundcolor=\color{backcolour},
  commentstyle=\color{codegreen},
 keywordstyle=\color{magenta},
    numberstyle=\tiny\color{codegray},
    stringstyle=\color{codepurple},
    basicstyle=\small\ttfamily, 
    breakatwhitespace=false,         
    breaklines=true,                 
    captionpos=b,                    
    keepspaces=false,                 
    numbersep=5pt,                  
    showspaces=false,                
    showstringspaces=false,
    showtabs=false,                  
    tabsize=2,
    escapeinside=``
}
\title{Hallucinations in Code Change to Natural Language Generation: Prevalence and Evaluation of Detection Metrics}

\author{Chunhua Liu$^1$ \quad Hong Yi Lin$^1$ \quad Patanamon Thongtanunam$^1$ \\
        $^1$School of Computing and Information Systems, 
        The University of Melbourne\\
        \texttt{chunhua.liu1@unimelb.edu.au}\\
        \texttt{holin2@student.unimelb.edu.au}\\
        \texttt{patanamon.t@unimelb.edu.au} 
}

\begin{document}
\maketitle
\begin{abstract}
Language models have shown strong capabilities across a wide range of tasks in software engineering, such as code generation, yet they suffer from hallucinations. While hallucinations have been studied independently in natural language and code generation, their occurrence in tasks involving {code changes which have a structurally complex and context-dependent format of code remains largely unexplored.}
This paper presents the first comprehensive analysis of hallucinations in two critical tasks involving code {change} to natural language generation: commit message generation and code review comment generation. We quantify the prevalence of hallucinations in recent language models and explore a range of metric-based approaches to automatically detect them. Our findings reveal that approximately 50\% of generated code reviews and 20\% of generated commit messages contain hallucinations. 
Whilst commonly used metrics are weak detectors on their own, 
combining multiple metrics substantially improves performance. 
Notably, model confidence and feature attribution metrics effectively contribute to hallucination detection, showing promise for inference-time detection.\footnote{All code and data will be released upon acceptance.} 


\end{abstract}
\section{Introduction}


AI-based software engineering tools are becoming increasingly ubiquitous due to their potential to improve developer productivity \cite{Jain2022Jigsaw, Fan2023LLMSE, Hou2024LLMSE}.
While such tools can accelerate software development, their reliance on underlying language models exposes the risk of hallucination—the phenomenon where models generate outputs that are inconsistent with their inputs or fabricate non-existent information \cite{ji2023survey, Huang2025Survey}.
Such behavior may decrease developer productivity or even mislead junior developers \cite{ferino2025junior}, allowing errors to propagate through to the software.
Although prior research has focused on the effects of hallucination during code generation \cite{liu2024exploring, tian2024codehalu, agarwal2024codemirage}, these effects remain largely unexplored in generation tasks involving code changes.
Unlike complete code files, code changes present snippets of both the old and new versions simultaneously, which could potentially amplify hallucinations due to the model's need to process and reason about multiple code states with partial context.


Indeed, code changes commonly used in the software engineering workflows~\cite{Taoetal2012, Lucaetal2023DiffSearch}. Recent work also leveraged code changes as primary inputs of language models for automated software engineering tasks such as code reviews~\cite{Li2022CodeReviewer, Lin2023CCT5}.
Given the increasing use of code changes in generation tasks, there is a need to understand the prevalence and effectiveness of the current detection metrics. The fragmented and context-dependent nature of code changes may increase hallucination risk and hinder detection.

In this paper, we present a comprehensive study of hallucinations in code {change} to natural language (\ccnl) generation tasks. We focus on two key tasks: (1) automated commit message generation, which aids developers in documenting what and why code was changed, and (2) automated code review generation, which assists reviewers in identifying potential issues in code changes and suggesting improvements. 
To systematically analyze hallucination in \ccnl, we first develop a hallucination annotation workflow specific to the \ccnl context based on the outputs from task-specific models. We then empirically evaluate the effectiveness of various metric-based approaches for automatically detecting these hallucinations.
In particular, we examine both reference-based metrics (which compare against human-written references) and reference-free metrics (without the references). 

Our findings reveal the severity of the hallucination problem in \ccnl tasks. 
We found that nearly 50\% of model-generated code reviews and 20\% of generated commit messages contain hallucinations. 
The three predominant categories of hallucinations are input inconsistency (where the generated NL is inconsistent with the code change), logic inconsistency (where the NL contains internally contradictory reasoning), and intention violation (where the generation fails for the specific task, e.g., it is not a review comment for code review but just a summary of the code change). 
Furthermore, we demonstrate that individual metrics for hallucination detection perform only marginally better than random chance (56.6\% ROC-AUC for code review and 61.7\% for commit messages). 
However, combining multiple metrics yields substantial improvements (69.1\% and 75.3\% respectively). Notably, reference-free metrics show promising results comparable to using all available metrics, suggesting the feasibility of detecting hallucinations without ground truth references.

This work makes three primary contributions: (1) the first systematic characterization of hallucinations in code {change} to natural language  tasks, revealing the severity and patterns of the problem; (2) a comprehensive evaluation of automatic hallucination detection methods, demonstrating that combining multiple metrics significantly improves detection capability; and 
(3) identification of key reference-free metrics (model confidence and attribution scores) that effectively predict hallucinations, facilitating real-time detection in production environments without requiring reference text. 


\section{Related Work}

\paragraph{Hallucination in Natural Language Generation}
Initially, \citet{maynez-etal-2020-faithfulness} categorized hallucinations in summarization into two types: \textbf{intrinsic} hallucinations (where models misinterpret information present in the input, generating content that contradicts the source document) and \textbf{extrinsic} hallucinations (where models forge information absent from the input that cannot be verified using available information).  
Recently, \citet{Huang2025Survey} identified three subcategories of intrinsic hallucinations in LLMs: instruction-inconsistent (outputs are not consistent with the instruction), logic inconsistency (output itself exhibits internal logical contradictions), and context inconsistency (outputs are not consistent with the provided input context). \citet{Huang2025Survey} further refined these factual hallucinations by distinguishing between factual contradiction (outputs that can be grounded but contradict real-world knowledge) and factual fabrication (outputs that are completely made up with no basis in reality or verifiable facts). 
Research on hallucination in code generation tasks also grounds hallucination types based on these categories~\cite{liu2024exploring}.
This taxonomy aligns closely with our \ccnl tasks and serves as a foundation to determine the hallucination types in Section~\ref{ssec:hallucination_taxonomy}. 

\paragraph{Hallucination in Code to Natural Language Generation} 
Different from hallucination research in natural language to code generation, which primarily focuses on incorrect code generations e.g., dead/unreachable code, syntactic incorrectness~\cite{liu2024exploring, agarwal2024codemirage}, hallucination in code to natural language generation focuses on natural language utterances that are incorrect with respect to the code/task at hand.
Whilst many hallucinations in code generation can be verified by static analysis and execution~\cite{tian2024codehalu}, these solutions are not applicable for natural language outputs. Recent work examined hallucination in code-to-natural language tasks~\cite{Zhang2024Detecting,maharaj2024etf,kang2024Identifying}. 
However, they primarily focus on compilable code implementations (e.g., the full body of a method). 
For example, \citet{maharaj2024etf} studied entity-level hallucination in code summarization, where the input consists of a method-level function containing adequate contextual information. 
Yet, other code-to-natural language tasks involving snippets of code changes remain largely overlooked, despite their common use in real-world scenarios like commit message generation and code review {\cite{Lin2023CCT5}}.
Moreover,  due to the technical constraints of long-context modeling, snippets of code changes are often used as inputs for generation tasks instead of the complete code context~\cite{Lu2025Towards, Berabi2024DeepCode}.
The fragmented, context-dependent nature of code changes may increase hallucination risk and hinder detection, motivating our investigation into their prevalence and the effectiveness of existing metrics.

\paragraph{Automatic Hallucination Detection} Automatic hallucination detection methods fall into two broad categories: reference-based and reference-free. \textit{Reference-based} metrics use ground truth to gauge the quality of the generated outputs, using this quality as an estimation of hallucination. This includes lexical overlap such as BLEU \cite{papineni-etal-2002-bleu}, which evaluates n-gram similarity between generated and reference texts. This is widely used in both Code2NL and NL2NL tasks \cite{liu2018neural, tufano2021towards,Li2022CodeReviewer,liu2025too}. 
More advanced metrics use Natural Language Inference (NLI): the model output is treated as a ``hypothesis'' to be validated against the reference. An entailment classifier labels output as entailment or contradiction, which maps to faithful or hallucinated content \cite{manakul-etal-2023-selfcheckgpt, elaraby2023halo,hu2024refchecker, valentin2024cost}. \textit{Reference-free} methods operate in many open-ended generation settings, where a reference is unavailable, by analyzing internal model behaviors and input-output relationships. One family of approaches estimates uncertainty inside models during generation \cite{guerreiro-etal-2023-looking,huang2023look}, with hallucinations typically exhibiting lower confidence in probability distributions and higher entropy. Another promising line is feature attribution techniques \cite{tang2022reducing,chen2025attributive}, which examine how inputs influence outputs, e.g., when a model hallucinates, its attention patterns or hidden states behave anomalously.
While these metrics have been used to detect hallucinations in various NL2NL tasks, such as machine translation and question answering
\cite{guerreiro-etal-2023-looking,dale-etal-2023-detecting}, their capabilities in \ccnl tasks remain unknown.





\section{Study Design}

\subsection{Research Questions}
\textbf{RQ1: To what extent do task-specific language models hallucinate in code change to natural language tasks?}
Prior work on hallucination in software engineering has focused on code generation, which can be verified deterministically. 
However, little attention has been paid to hallucinations in \ccnl generation tasks, such as code review comment generation and commit message generation. 


\noindent\textbf{RQ2: How effectively can existing hallucination detection methods perform on code change to natural language tasks?}
While prior work in NLP have developed various methods \cite{dale-etal-2023-detecting, Huang2025Survey, ji2023survey} to detect hallucinations in natural language generation, their applicability to the bi-modal scenario of \ccnl remains unknown. 
Effective detection in such contexts requires an understanding of the semantics behind both code, natural language, and their interaction.

\subsection{Hallucination Annotation Workflow}
\label{ssec:hallucination_taxonomy}

\begin{figure}[!t]
    \centering
    \includegraphics[width=0.85\linewidth]{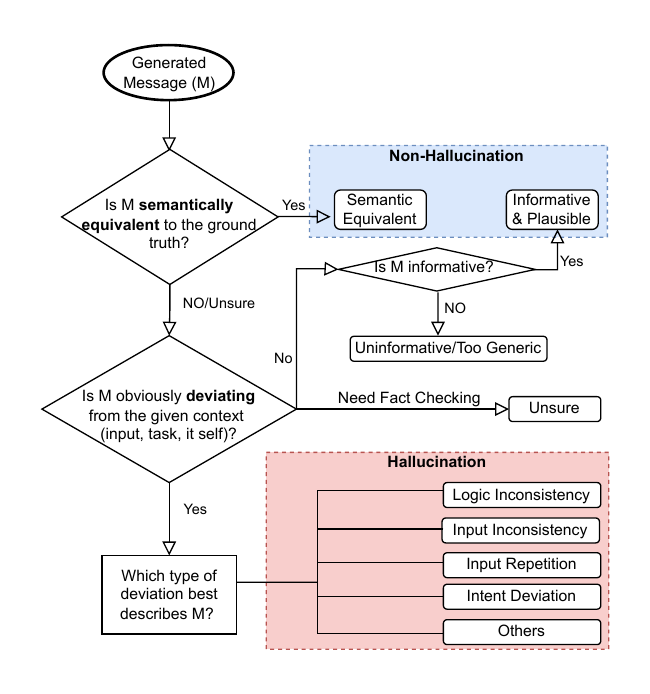}
    \caption{Hallucination Annotation Flowchart}
    \label{fig:hallucination_taxonomy_drawio}
\end{figure}
Since no existing work addresses hallucinations in the \ccnl context, we developed a decision-tree-based hallucination detection workflow by adapting taxonomies from both code generation \cite{liu2024exploring} and natural language hallucination \cite{Huang2025Survey}. 
Our workflow\footnote{See Appendix~\ref{sec:apdx:hallucination_annotation} for definition and annotation guidelines.} (see Figure~\ref{fig:hallucination_taxonomy_drawio}) evaluates a generated NL as follows:

\textbf{Semantic Equivalence.} We first determine whether the generated NL is semantically equivalent to the ground truth (i.e., conveying the same intent with similar framing and emphasis). If equivalent, the output is classified as non-hallucination. 

\textbf{Contextual Faithfulness.} For semantically non-equivalent outputs, we assess whether the NL deviates from the context (source code, task specification, and generated text itself). 
Non-deviating outputs are classified as either \textit{Informative \& Plausible} (valid alternatives) or \textit{Uninformative} (truisms).

\textbf{Hallucination Type Classification.} When context deviation exists, we categorize the hallucination into five types:\footnote{Examples are provided in Appendix \ref{apdx:ssec:hallucination_examples}.} 1) \textit{Input Inconsistency}, where the generation conflicts with the source code, e.g., pointing out a non-existent issue in code review or speculating intent that contradicts the code change in commit messages; 
2) \textit{Logic Inconsistency}, where the generation is internally illogical, independent of the input;
3) \textit{Input Repetition}, where the generation directly copies from the input; 
4) \textit{Intent Deviation}, where the generation deviates from the task's goal, e.g., not identifying issues in a code review or not explaining the code change in a commit message; and 
5) \textit{Others} for cases that are not covered by the above types. 
Cases requiring additional project specific fact-checking are labeled as \textit{Unsure}.


\subsection{Datasets and \ccnl Generation}
\textbf{Datasets.} We choose the widely used CodeReviewer \cite{Li2022CodeReviewer} dataset for code review comment generation and CommitBench \cite{schall2024commitbench} for commit message generation.  
The CodeReviewer corpus contains code diff and natural language review pairs, across 9 popular programming languages and over 1k GitHub projects. 
It includes 118k training, 10k validation, and 10K testing examples. 
CommitBench contains code diffs paired with natural language commit messages, spanning over 72k GitHub repositories and 6 programming languages. It includes 1.16 million training examples and 250k examples each for validation and testing.
While related, the two tasks are different in nature—commit messages are primarily descriptive, whereas code reviews require deeper reasoning about functional correctness 
and potential impacts across the codebase.

\textbf{Models.} 
To analyze hallucination behaviors, we conduct experiments to select language models that are highly capable in both tasks.
This is determined by BLEU-4 results,
which is the most commonly used metric~\cite{Li2022CodeReviewer,schall2024commitbench}. We choose two recent LLM families (Qwen2.5 and Llama3.1)\footnote{These were the latest models at the time of experiment.} with varied model sizes for both direct prompting (7-8B, 70-72B) and task-specific fine-tuning (7-8B). 
We also fine-tune CCT5~\cite{Lin2023CCT5}, which is a 220M T5-based model pre-trained on 1.5M code change to commit message pairs. We used the original training data in two datasets to fine-tune the models.
We found that fine-tuned models performed the best for both tasks.\footnote{{See Appendix~\ref{sec:apdx:prompting_fine_tuning_models} for details on prompting and fine-tuning.}}
Table~\ref{tab:code_review_commitbench_bleu_fine_tune} (Overall columns) presents the experimental results. Thus, we select the three fine-tuned models to generate outputs for hallucination analysis in Sections~\ref{sec:rq1} and ~\ref{sec:rq2}.

\begin{table}[t]
\centering
\scriptsize
\begin{tabular}{llcccc}
\hline
\textbf{Model} & \multicolumn{2}{c}{\textbf{CodeReview}} &  \multicolumn{2}{c}{\textbf{CommitBench}} \\
~ & Overall & Sample & Overall & Sample \\ 
\hline
 Llama3.1-8B       & 5.28 & 5.25 & 15.06 & 15.29\\
Qwen2.5-7B   & 5.43 & 5.73 & 15.37 & 15.57 \\
CCT5         & \textbf{5.58} & \textbf{6.53} & \textbf{17.45} & \textbf{17.46} \\
\bottomrule
\end{tabular}
\caption{Performance (BLEU-4 in \%) of fine-tuned models on CodeReview and CommitBench benchmarks. 
}
\label{tab:code_review_commitbench_bleu_fine_tune}
\end{table}

\subsection{Hallucination Detection Methodology}
\label{ssec:hallu_detect_methods}
\begin{table*}[!t]
\centering
\scriptsize
\begin{tabular}{l|l|p{9cm}}
\hline
\textbf{Metric} &  \textbf{Type} & \textbf{Description} \\
\hline
BLEU-4 &  Lexical-Overlap & The n-gram overlap between the generation $y$ and reference $\hat{y}$.\\
\hline
Entailment &   NLI & The probability that a NLI classifier predicts $\hat{y}$ entails $y$. We used nli-deberta-v3\footnote{https://huggingface.co/cross-encoder/nli-deberta-v3-base} as the classifier. \\
\hline
Similarity &   Similarity & The embedding-based cosine similarity between the generation $y$ and source code $x$. We used three embeding models: codebert-base\footnote{https://huggingface.co/microsoft/codebert-base}, codet5p-220m-bimodal\footnote{https://huggingface.co/Salesforce/codet5p-220m-bimodal}, and codet5p-770m\footnote{https://huggingface.co/Salesforce/codet5p-770m}. \\
\hline
SeqLogProb &  Uncertainty & The average negative log-probability of the generated tokens in $y$ as assigned by a language model $M$. \\
\hline
SeqLogit &  Uncertainty & The average raw logit score (pre-Softmax) of the generated tokens in $y$ from a model $M$.  \\
\hline
SeqEntropy &  Uncertainty & The average entropy of the generated tokens in $y$ from a model $M$.  \\
\hline
Source Attribution &   Feature Attribution & The average of the maximum attribution scores from source tokens to each generated token in $y$ (i.e., $\frac{1}{T}\sum_{t=1}^{T} \max_{i \in [1,N]} A_{i,t}$, where $A_{i,t} = x_i \times \frac{\partial y_t}{\partial x_i}$ is the importance of $x_i$ to $y_t$ from a model $M$). A higher score represents source contributes more strongly to $y$.  \\
\hline
Target Attribution &   Feature Attribution & The average of the maximum attribution scores from previously generated tokens ($y_1, \dots, y_{t-1}$) to each current token $y_t$. A higher score represents the reliance on previously generated tokens. \\
\hline
Changed Attribution &   Feature Attribution & The average of the maximum attribution scores from source tokens that are changed (in +, - lines) to each generated token in $y$.  A high score represents changed tokens contributes strongly to $y$.\\
\hline
Unchanged Attribution &  Feature Attribution & The average of the maximum attribution scores from source tokens that are unchanged to each generated token in $y$. A high score represents unchanged snippets in source contributes strongly to $y$. \\
\hline
\end{tabular}
\caption{Descriptions of hallucination detection metrics, including into \textit{reference-based} (BLEU-4 and Entailment) and \textit{reference-free} (all others). For uncertainty and feature attribution, the model $M$ $\in \{ \text{LLaMA3.1-8B}, \text{Qwen2.5-7B}, \text{and CCT5}\}$. 
We apply both self-attribution (generator attributes its own output)
and cross-attribution (external model attributes generator's output).
See Appendix~\ref{ssec:apdx:hallu_detection_methods} for a detailed description.}
\label{tab:hallucination_detection_metrics_description}
\end{table*}


We use both reference-based and reference-free hallucination detection approaches: the former for model development where the ground truth is available, and the latter for real-world deployment where references are unavailable. Table~\ref{tab:hallucination_detection_metrics_description} presents a summary of the metrics we used, 
including two types of reference-based (BLEU-4 and NLI), and three types of reference-free (similarity, uncertainty, and feature-attribution). 
Uncertainty and feature-attribution metrics are calculated with either LLaMA3.1-8B-Instruct~\cite{grattafiori2024llama},
Qwen2.5-7B-Instruct~\cite{qwen2025qwen25technicalreport} or CCT5~\cite{Lin2023CCT5}.
Due to space limitations, detailed descriptions and formulas are provided in {Appendix~\ref{ssec:apdx:hallu_detection_methods}}. 
In total, 26 unique methods were considered:  
2 reference-based metrics + 3 similarity scores + 3 models × 7 feature attribution and uncertainty metrics.

\section{To what extent do task-specific language models hallucinate in \ccnl tasks?}
\label{sec:rq1}

To address RQ1, we manually categorize the messages generated by the three fine-tuned models into our \ccnl hallucination annotation workflow introduced in Section~\ref{ssec:hallucination_taxonomy} to identify the presence and types of hallucinations. Using the annotated samples, we further analyze the overall prevalence of hallucinations and their distributional patterns across models and two datasets. 


\begin{table*}[htbp]
\centering
\scriptsize
\begin{tabular}{ll|ccc|ccc}
\toprule
\textbf{Category} &\multirow{2}{*}{\textbf{Type}} & \multicolumn{3}{c|}{\textbf{CodeReviewer}} & \multicolumn{3}{c}{\textbf{CommitBench}} \\
\cmidrule{3-8}
 & & CCT5 & Llama3.1 & Qwen2.5 & CCT5 & Llama3.1 & Qwen2.5 \\
\midrule 
\multirow{2}{*}{Non-Hallucination}& Semantic\_Equivalent & 1.5 & 1.1 & 1.5 & 11.2 & 12.3 & 16.4 \\
& Informative & 9.5 & 9.8 & 8.7 & 48.1 & 42.5 & 44.4 \\ \midrule
Uninformative & Uninformative & 20.1 & 1.5 & 3.8 & 15.7 & 7.1 & 9.7 \\\midrule 
Unsure & Unsure & 22.0 & 41.3 & 43.2 & 5.6 & 16.4 & 15.3 \\\midrule  
\multirow{5}{*}{Hallucination}& Input\_Inconsistency & 26.5 & 23.9 & 24.6 & 17.2 & 19.8 & 13.1 \\
& Input\_Repetition & 4.2 & 0.0 & 0.0 & 0.0 & 0.7 & 0.7 \\
& Intent\_Deviation & 0.8 & 17.4 & 15.9 & 0.4 & 0.4 & 0.0 \\
& Logic\_Inconsistency & 14.0 & 4.5 & 1.9 & 1.9 & 0.7 & 0.4 \\
& Others & 1.5 & 0.4 & 0.4 & 0.0 & 0.0 & 0.0 \\
\midrule 
Total Hallucination	& ~ & 47.0	& 46.2 &	42.8	& 19.5 &	21.6 &	14.2 \\ 
\bottomrule
\end{tabular}
\caption{The distribution (percentage) of hallucination categories and types for annotated samples. The Category column is the high-level category in Figure~\ref{fig:hallucination_taxonomy_drawio}. The ``Total Hallucination'' is the sum of the four hallucination types.
}
\label{tab:rq1-manual-labeling}
\end{table*}
\subsection{Manual Annotation}
We selected the top 3 fine-tuned models (lama3.1-8B, Qwen2.5-7B, and CCT5) to generate messages in the test set. 
To address RQ1, we manually labeled a subset of 
samples that were randomly selected from the test set of each task, constituting a statistically significant sample size with a confidence level of 90\% and a margin of error of ±5\%. This results in 264 samples for CodeReviewer comments and 268 samples for CommitBench. In total, we annotated 1,596 samples, including 264 $\times$ 3 model outputs for CodeReviewer comments and 268 $\times$ 3 for CommitBench messages.

Two annotators (authors of the paper) with 5+ years of experience in computer science and software engineering annotated all samples. We conducted two pilot rounds (150 samples each) to refine the taxonomy and guidelines. Cohen’s $\kappa$ improved from 0.36/0.30 (CodeReviewer/CommitBench) in the first round to 0.56/0.38 in the second. Final disagreements were resolved through discussion, achieving near-perfect agreement ($\kappa = 0.98$ / $0.96$). The annotators then divided the remaining samples (half-half), cross-examining each other's work to ensure consistent labeling.

\subsection{Hallucination Prevalence and Patterns}



Table~\ref{tab:rq1-manual-labeling} shows that hallucination rates vary significantly across tasks. For the code review task, all models exhibit high hallucination rates ranging from 42.8\% to 47.0\%.
Surprisingly, although CCT5 achieves the highest BLEU score on the CodeReviewer dataset among the three models (Table~\ref{tab:code_review_commitbench_bleu}), it also exhibits the highest hallucination rate at 47.0\%. This highlights the risk of hallucinations even in models with strong BLEU performance.
On the other hand, the commit message generation task has a lower hallucination rate than code review (14.2\% to 21.6\%), where Qwen2.5 has the lowest rate at 14.2\%. 
This may be because code review is more challenging than commit message generation, as it requires identifying problems and providing specific feedback beyond what is directly observable in the code changes. 
Such added complexity might lead to increased hallucination behavior.





The overall distribution of hallucination types varies between tasks.
Notably, the Input Inconsistency emerges as the dominant hallucination type for both tasks. This suggests that models frequently generate messages that contradict or misrepresent the actual code changes. 
One frequent issue in code review is that the generated messages tend to fabricate non-existent code tokens. 
For example, CCT5 suggests \textit{``I think this should be \lstinline|orderPath| instead of \lstinline|orderPathKey|''}.
However, \lstinline|orderPathKey| does not appear in the code change:\footnote{The full code context in provided in Appendix ~\ref{apdx:patch_examples_codereview}.} \lstinline|+~public static final String ORDER\_PATH = "orderPath";|
This suggests that the model does not fully understand the meaning of newly introduced code. In the commit message task, models also often misunderstand the code changes. For example, the generated message \textit{``nomad: fix peers.json recovery for protocol version 3''} misrepresents the change, which actually adds support for Nomad versions \textbf{below 3}, as indicated by the code line \lstinline|+ if s.config.RaftConfig.ProtocolVersion < 3 {|.\footnote{The code patch is  provided in Appendix ~\ref{apdx:patch_examples_commitbench}.} 

Intent deviation and logic inconsistency appear as another two pronounced hallucination types in the code review task, but they are rare in the commit message generation, suggesting that commit message generation models generate messages that better align with the task and suffer less logic inconsistency. Interestingly, we observe many cases where the generated review comment reads more like a commit message—for example, \textit{``This is a temporary fix.''}, which describes the code change rather than providing a review.


Different models exhibit different type of hallucinations. CCT5, which is the specialized fine-tuned model demonstrates higher logic inconsistencies (14.0\% in CodeReviewer) but significantly lower intent deviation (0.8\%) than general-purpose LLMs. 
On the other hand, larger models (Llama3.1, Qwen2.5) frequently have intent deviation ($\geq$ 15.9\% average) but fewer logic inconsistencies ($\leq$4.5\%). This pattern likely reflects the difference between specialized and general-purpose pretraining.
Despite fine-tuning, general models retain broad task knowledge from pretraining, which can lead them to apply reasoning patterns from unrelated tasks—resulting in higher intent deviation.




\section{How well do existing metrics detect hallucinations in \ccnl tasks?} 
\label{sec:rq2}


RQ1 showed that models often exhibit  hallucinations and misinterpretations of code changes. 
In RQ2, we examine how effective automated approaches are at detecting these hallucinations in code review and commit message generation.
Using our manually annotated dataset, we evaluate both reference-based and reference-free metrics described in Section~\ref{ssec:hallu_detect_methods}.
Our goal is to assess how well existing metrics detect hallucinations in Code-to-NL tasks, particularly for code changes. We evaluate both individual metrics and combinations of complementary ones to determine whether they can  approximate human judgment.

We use ROC-AUC to evaluate the hallucination detection capability of each metric. 
The positive class is the  hallucination samples that we annotated. The negative class is the non-Hallucination samples.
A ROC-AUC score of 1 indicates perfect discrimination between hallucinated and non-hallucinated cases, while a score of 0.5 suggests no discriminatory power equivalent to random guessing.
For individual metrics, we calculate the ROC-AUC to assess discrimination power.\footnote{The point-biserial correlation confirms a similar trend between metric scores and hallucination labels. Detailed results are provided in Appendix~\ref{ssec:apdx:point_biserial_correlation}.} 
To combine metrics, we use logistic regression and evaluate its performance using accuracy and ROC-AUC.

\begin{figure*}
    \centering
    \includegraphics[width=\linewidth]{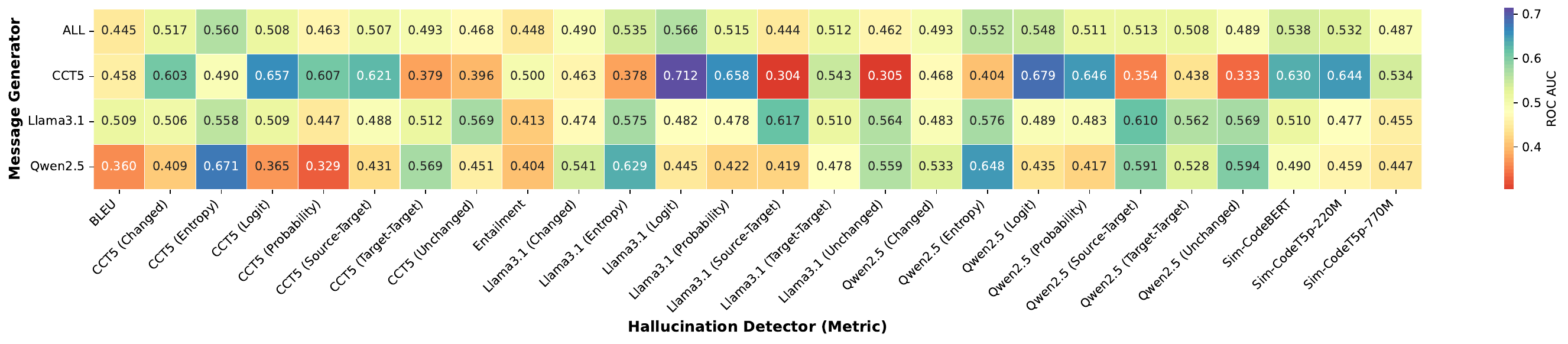}
    \caption{ROC-AUC Scores of Metrics for Hallucination Detection Across Generators on CodeReviewer. The ALL row represents the generator-agnostic result, using all outputs from CCT5, Llama3.1, and Qwen2.5.
The remaining rows show performance in the generator-specific result, based on outputs from each model individually.}
    \label{fig:individual_metrics_heatmap_roc_auc_codereviewer}
\end{figure*}

\begin{figure*}
    \centering
    \includegraphics[width=\linewidth]{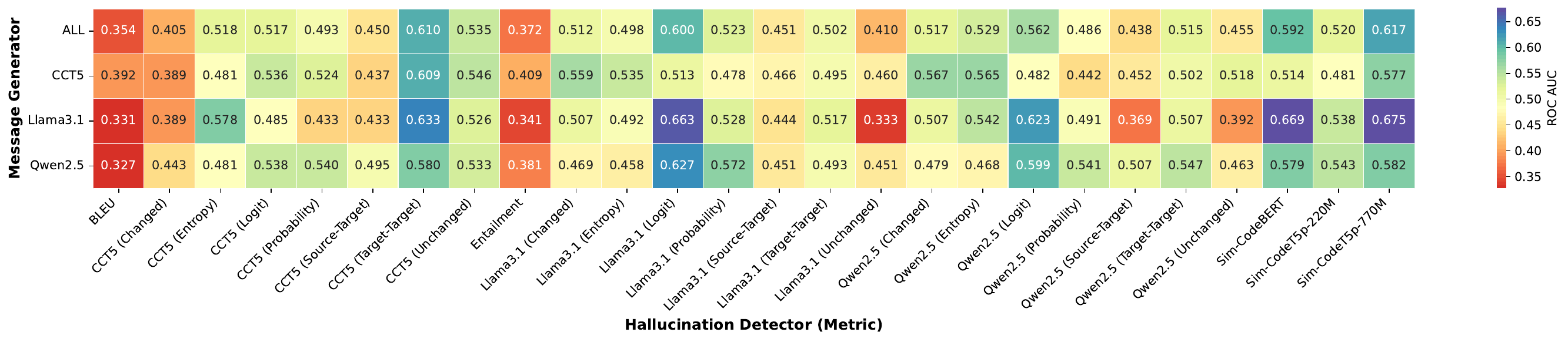}
    \caption{ROC-AUC Scores of Metrics for Hallucination Detection Across Generators on CommitBench.}
    \label{fig:individual_metrics_heatmap_roc_auc_commitbench}
\end{figure*}

\subsection{How do individual metrics perform in detecting hallucinations?}
\label{sec:individual_metrics}

\textit{Metric Effectiveness.} Based on the the generator-agnostic results, the current metrics achieve modest ROC-AUC scores ranging from 0.538--0.566 on CodeReviewer   and 0.562--0.617 on CommitBench (see Figures \ref{fig:individual_metrics_heatmap_roc_auc_codereviewer} and  \ref{fig:individual_metrics_heatmap_roc_auc_commitbench}).
Based on the generator-specific results, hallucinations in CCT5 are more detectable on the CodeReviewer dataset (ROC-AUC 0.65-0.71), while hallucinations in Llama3.1 are most detectable on the CommitBench dataset (ROC-AUC 0.62-0.68).
This suggests that the effectiveness on hallucination detection of the metrics may vary across generation models and datasets.

Table \ref{tab:experimental_results_on_logistic_regression} shows the metrics with the highest ROC-AUC scores in each studied dataset. 
In addition, we observe that on CodeReviewer, uncertainty-based metrics (logit and entropy) perform best, while embedding similarity and reference-based metrics are best on CommitBench. 
Nonetheless, the ROC-AUC scores suggest the limited effectiveness of current metrics on hallucination detection, which are slightly better than random guessing, highlighting the challenges of automated hallucination detection in these tasks.



\begin{figure}[!t]
  \centering
  \begin{subfigure}[t]{0.23\textwidth}
    \centering
    \includegraphics[width=\linewidth]{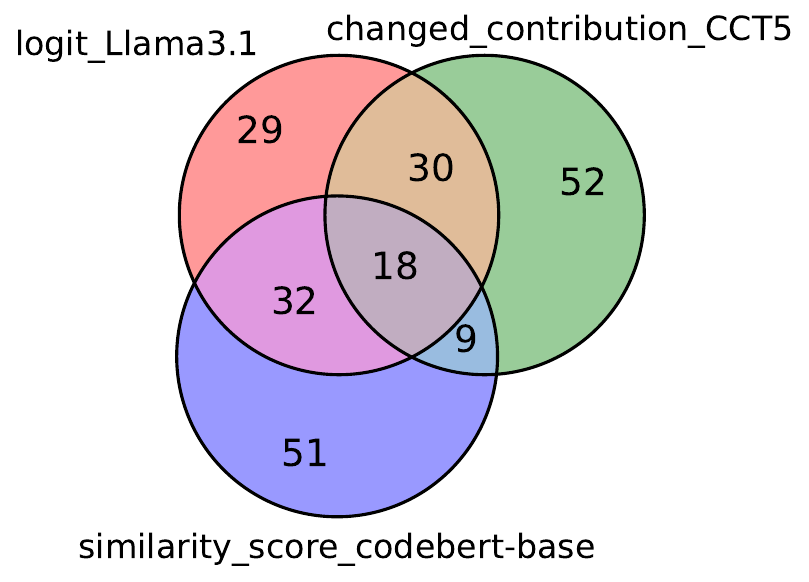}
  \end{subfigure}
  \hfill 
  \begin{subfigure}[t]{0.23\textwidth}
    \centering
    \includegraphics[width=\linewidth]{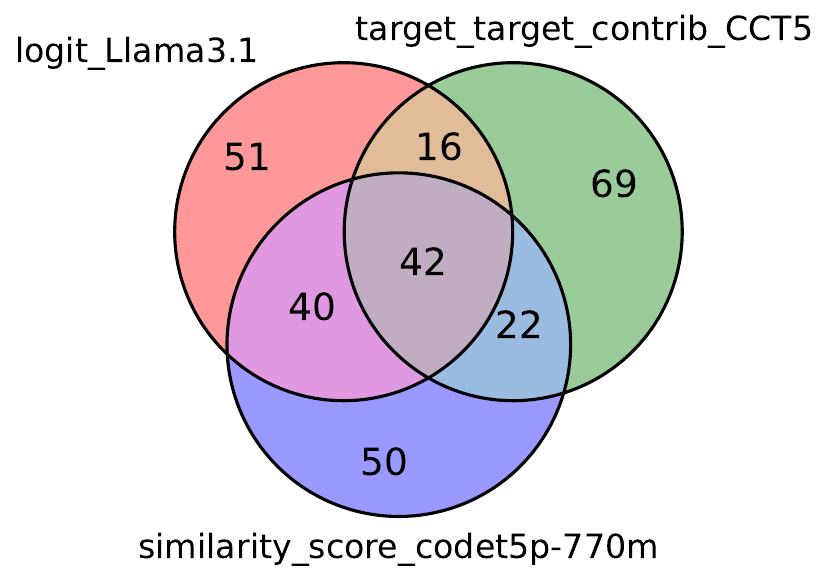}
    
  \end{subfigure}
  \caption{Top 3 individual metrics complement to each other on CodeReviewer (left) and CommitBench (right)}
  \label{fig:venn_diagram_three_metrics}
\end{figure}

\textit{Metric Complementarity.}
Different metrics may capture distinct aspects of hallucinations, potentially flagging different instances.
To assess this, we selected the three highest-performing metrics based on ROC-AUC and examined their top 25\% ranked samples (see the analysis details in Appendix~\ref{apdx:sec:metric_complementarity}).
Figure~\ref{fig:venn_diagram_three_metrics} shows small overlap in the top 25\% samples ranked by these three metrics, indicating these metrics flag different instances as hallucinated.
This highlights the potential complementarity between metrics.

\subsection{Can combining multiple metrics enhance the accuracy of hallucination detection?}
\label{ssec:combien_multiple_metrics}

The results in section \ref{sec:individual_metrics} highlight the potential complementarity between metrics. 
Thus, we explore whether combining them can improve performance. Prior work \cite{Snyderetal2024-Early} also shows that combining multiple signals improves hallucination detection in question-answering tasks.
To analyze the descrimination power of combined metrics for hallucination detection, we use a logistic regression model fitted to our annotated samples. For each generation task, we combine all samples from the three models, resulting in 440 samples for CodeReviewer and 717 samples for CommitBench. 

To understand the capability of different types of metrics, we build three logistic regression models using: 1) all metrics, 2) reference-based metrics only, and 3) reference-free metrics only.
Since some metrics may capture similar signals or redundant, leading to multicollinearity and overfitting, we use the Akaike Information Criterion (AIC)~\cite{Akaike1974AIC} to identify metrics that meaningfully contribute to the prediction.
Then, we use the selected metrics as features to fit the logistic regression model and analyze the coefficients to identify which metrics are most important for hallucination detection.


\begin{table}[!t]
    \footnotesize
    \centering
    \resizebox{0.48\textwidth}{!}{
    \begin{tabular}{l|cc | cc }
        ~ & \multicolumn{2}{c|}{\textbf{CodeReviewer}} & \multicolumn{2}{c}{\textbf{CommitBench}} \\ \toprule
        \textbf{Type} & \textbf{Acc} & \textbf{AUC}  & \textbf{Acc} & \textbf{AUC} \\ \midrule
        \multicolumn{5}{c}{Top Performing Individual Metrics} \\ \midrule
        logit\_Llama3.1 & - & 0.57 & - & 0.60 \\  
        Sim-CodeT5p-770M & - & 0.48 & - & 0.62 \\
        Sim-Codebase & - & 0.54& - & 0.59\\ 
        changed\_contrib\_CCT5 & - & 0.52 & - & 0.41 \\ 
        target\_target\_contrib\_CCT5 & - & 0.49 & - & 0.61\\
        \midrule 
        \multicolumn{5}{c}{Multiple Metrics on Logistic Regression} \\ \midrule 
        Reference-based &   81.6 & 0.59 &  76.0 & 0.68 \\ 
        Reference-free &    81.6 &  0.66    & 78.9 & 0.75 \\
        ALL & \textbf{82.7} & \textbf{0.69} & \textbf{77.8} & \textbf{0.75}  \\
        \bottomrule
    \end{tabular}
    }
    \caption{Logic regression results (Acc (\%) and AUC) on hallucination prediction using multiple metrics.}
    \label{tab:experimental_results_on_logistic_regression}
\end{table}


Table~\ref{tab:experimental_results_on_logistic_regression} shows the logistic regression results. Combining multiple metrics substantially improves ROC-AUC scores for hallucination detection on both datasets, compared to individual metrics alone. For CodeReviewer, the ROC-AUC increased from the best individual score of 0.57 (logit\_Llama3.1) to 0.69 when using all metrics. For CommitBench, it improved from 0.62 (similarity\_score\_codet5p-770m) to 0.75.
Surprisingly, using reference-free metrics alone achieved ROC-AUC scores close to that of using all metrics.
In contrast, reference-based metrics achieved lower performance, possibly because they are fewer in number or inherently less predictive.
This highlights a potential benefit of hallucination detections in these \ccnl tasks without ground-truth.


\begin{table}[!t]
\scriptsize
\centering
\begin{tabular}{ll|c|c}
\toprule
\textbf{Type} & \textbf{Metric} & \textbf{|Coef|} & \textbf{Sign} \\
\midrule 
Uncertainty & logit\_Llama3.1 & 6.00$^*$ & + \\
Uncertainty & entropy\_Qwen2.5 & 3.33$^*$ & + \\
Attribution & source\_target\_Qwen2.5 & 2.83$^*$ & + \\
Attribution& source\_target\_Llama3.1 & 2.78$^*$ & - \\
N-gram &BLEU & 1.94$^*$ & - \\
\bottomrule 
\end{tabular}
\caption{\label{tab:rq2_logist_regression_codereviewer}Top-5 important features on predicting hallucinations (vs. non-hallucinations)  in CodeReviewer. $*$ indicates the coef is significant ($p<0.05$).}
\end{table}


\begin{table}[!t]
    \scriptsize
    \centering
    \begin{tabular}{ll|cc}
    \toprule
    \textbf{Type} & \textbf{Metric} & \textbf{|Coef|} & \textbf{Sign} \\
    \midrule 
    Uncertainty & logit\_Llama3.1 & 6.86$^*$ & + \\
    Uncertainty & logit\_Qwen2.5 & 5.93$^*$ & - \\
    Attribution & changed\_CCT5 & 4.71$^*$ & - \\
    N-gram & BLEU & 3.49$^*$ & - \\
    Similarity &similarity\_score\_codebert & 2.41$^*$ & + \\
    \bottomrule 
    \end{tabular}
     \caption{\label{tab:rq2_logist_regression_commitbench}
   Top-5 importance features on predicting hallucinations (vs. non-hallucinations) in CommitBench. 
   }
\end{table}


Tables~\ref{tab:rq2_logist_regression_codereviewer} and~\ref{tab:rq2_logist_regression_commitbench} present the most important features along with their coefficients. The SeqLogit calculated with Llama3.1 (Logit\_Llama3.1) emerges as the most important feature for both tasks. Uncertainty metrics from Llama3.1 and Qwen2.5 consistently appear among the top features, demonstrating strong predictive power.For CommitBench dataset, the $-$ coefficient in Qwen2.5 aligns with prior findings, i.e., when not hallucinating, a model is more confident~\cite{dale-etal-2023-detecting}.
On the other hand, the $+$ coefficient in Llama3.1 could be due to its overconfidence as the distribution of logit in Llama3.1 is skewed towards high scores in {CommitBench}
(See Appendix~\ref{apdx:ssec:lr_opposite_sign_coef}). Feature attribution metrics rank next in predictive strength, indicating that hallucinations can be detected by analyzing how models utilize source code during generation.
\begin{figure}[!t]
    \centering
    \includegraphics[width=\linewidth]{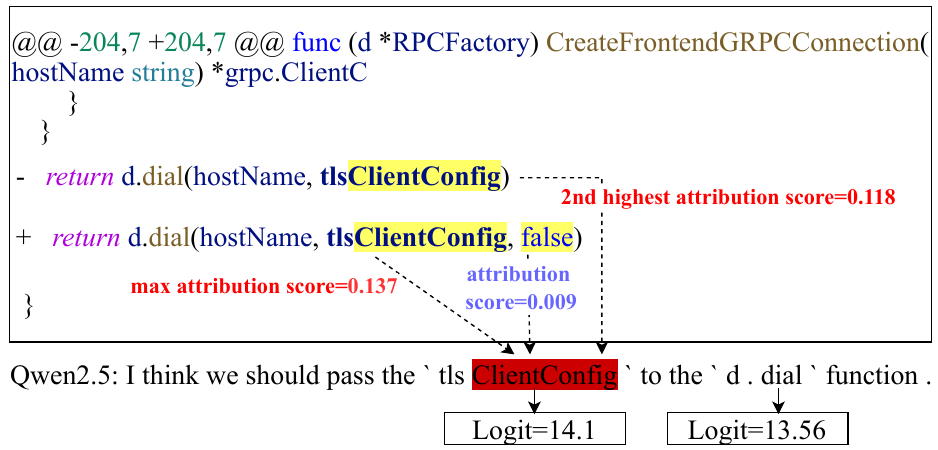}
    \caption{An example of feature attribution on a hallucinated code review comment generated by Qwen2.5. Attribution model: Llama3.1.}
    \label{fig:case_study1}
\end{figure}

Figure~\ref{fig:case_study1} presents an example generated for code review.\footnote{See an example of commit message in Appendix Figure~\ref{fig:comit_bench_case_study}.}
The generated review suggests passing a parameter that is already being passed in both old and new code, while ignoring the actual code change. 
This hallucinated generation has high logit and high attribution from source code. Particularly, the generated tokens appearing in the input context have high confidence based on elevated logit values.
For example, based on uncertainty calculated with Llama3.1, particularly API method names like \lstinline|tlsClientConfig| and \lstinline|dial| have logit values of 14.1 and 13.6. 
However, based on the attribution scores,  critical changes (i.e., the addition of the ``\lstinline|false|'' parameter) that should be the primary focus of the review has  minimal contribution to the generation. 
Instead, these common tokens like \lstinline|tlsClientConfig| have large attribution scores, meaning that they contributing significantly to the generation.



For non-hallucinations, we observed that the correct input in the code changes contributes significantly to the relevant generation compared to other code snippets (e.g., in the generated comment, ``\textit{Why is this needed?}'' the ``this'' token was mainly contributed by the changed line of code ``\textit{+ from databricks import koalas as ks}''). This indicates that the balance between the contribution from changed code and unchanged code is one important cause of hallucination in code review tasks.

\section{Conclusion}
Hallucinations are prevalent in \ccnl tasks, occurring in 50\% of code reviews and 20\% of commit messages.
We identify three common types—input/logic inconsistency, and intention violation.
Our findings show that individual metrics are insufficient for effective detection, while a multi-metric approach significantly improves performance, particularly combining model confidence and feature attribution.

\section{Limitations}

While our study advances the understanding of hallucination severity and automatic detection capabilities in \ccnl tasks, several limitations remain.

\paragraph{Dataset Size.} Despite using statistically representative samples from the test set, our annotated dataset is relatively small due to the significant effort required for manual annotation. To mitigate this limitation, we analyzed both model-specific and aggregated samples across models to increase effective sample sizes. 

\paragraph{Hallucination Granularity.} We primarily focused on instance-level (whole sequence) hallucination analysis to establish a foundational understanding of the phenomenon. Our feature attribution analysis showed promise for token-level hallucination detection, revealing cases where generation heavily relied on unchanged code snippets while ignoring critical changes. Future work should explore finer-grained token-level hallucination analysis with appropriate annotations and develop techniques for more precisely identifying hallucinations at different levels of granularity.




\paragraph{Model Recency and Coverage.} Due to cost constraints, we excluded commercial models (e.g., GPT-4o, Claude 3.7) from our analysis and focused on the latest open-source language models available at the time of our experiments. However, the landscape is evolving rapidly, with newer models such as LLaMA 4 and Qwen2.5-Coder emerging since our evaluation. As a result, our findings may not fully generalize to these newer or commercial models, or to different model families such as Gemini, which could exhibit different hallucination patterns in Code2NL tasks. Also, our study focuses on the hallucination in task-specific fine-tuned models since they perform better than zero-shot prompting.
The hallucination prevalence in zero-shot prompting may be different. 
Our work lays the foundation for future research in this space, highlighting the need for ongoing evaluation as models continue to evolve and diversify.



\bibliography{main.bib}

\appendix

\begin{table*}[!t]
    \centering
    \scriptsize
    \begin{tabular}{p{15cm}}
        \textbf{Type}:  Definition \\
        \midrule 
         \textbf{Semantic Equivalent (SE)}:  The generated message is semantically equivalent to the ground truth.  \begin{itemize}
            \item In code review, a semantically equivalent comment should share the same intentions regarding both the issues identified and the solutions proposed as in the ground truth. 
             \item In commit message, we should consider both the ``What'' and ``Why'' together to decide the semantic equivalence. Semantic equivalent commit messages should convey the same intents with similar framing and emphasis.
         \end{itemize}\\ \hline 
         \textbf{Not\_SE\_Informative}:  M is different from ground truth but it is informative for the task as hand. \begin{itemize}
             \item In code review, M is considered as informative if it points out a concern and/or provide suggestions for improvement. 
             \item For commit messages, M captures some aspects of the code change but may overlook certain points compared to the ground truth. For instance, \textit{`Add 'scheme' to sys path in ok\_test/scheme.py''} indicates where the change occurs but lacks the 'why.' In contrast, the ground-truth message \textit{Add 'scheme' to path to handle zip archive case''} provides (why) context on the purpose of the modification.
             Note (simple way): M must contain ``What'', but can be incomplete or slightly different from ground truth; ``Why'' can be missing.
         \end{itemize} \\  \hline 
         \textbf{Not\_SE\_Uninformative}: M is different from the ground truth and it doesn't provide useful information for the task at hand. \begin{itemize}
             \item In code review, M is considered uninformative if it merely seeks information to understand the code design or implementation choices, presents a general question without rationale, serves as self-justification for the code change, or acts as a compliment to the code. Note (simple way): if the What (issue) is missing, then it's not informative.
             \item In commit messages, vague and general wording fails to clearly communicate the specifics of the change, such as the `what' (the nature of the modification) and the `why' (the reason for the modification). For example, the message \textit{`Minor refactoring in VRaptor'} lacks detail about what parts were refactored and the intended impact of those changes, making it difficult for reviewers to understand the significance or context of the update. 
             Note (simple way): ``What'' is essential, it's uninformative if it lacks specifics of ``What''.
         \end{itemize} \\  \hline 
         \textbf{Unsure\_or\_Looks\_Applicable}:   M appears relevant to the context but needs further fact-checking, as its factual accuracy cannot be directly verified from the given context \begin{itemize}
             \item In code review, this can involve M using context such as historical background, rationale beyond the given input, or the need for fact-checking the provided solution. 
             \item In a commit message, the rationale for explaining the issue or objectives in M might need fact-checking.
         \end{itemize}
         \\ \hline 
        \textbf{Input Inconsistency} : M conflicts with the provided input. \begin{itemize}
         \item In code review, this means M points out an non-existent issue or provides a solution that is already exists in the code change or violates with programming commonsense. 
         \item In commit message, this means that M contains information that's not included in the code change, or misinterpret code change.
         \end{itemize}
          \\ \hline 
          \textbf{Logic Inconsistency}: M itself doesn't make logical sense.\\  \hline 
         \textbf{Context Repetition}:  M is completely or largely copied from the input. \\ \hline  
         \textbf{Intent Deviation}: M deviates with the goal of the task at hand: not providing a review in code review task or not providing a commit message that covers what is being changed and why it's being changed.\\  \hline 
         \textbf{Others}: This is used to capture any other types that's not covered in the above categories \\ 
         \bottomrule 
    \end{tabular}
    \caption{The definitions for each of the type in our annotation. M denotes the model generated message.}
    \label{tab:annotation_label_definition}
\end{table*}

\section{Hallucination Annotation }
\label{sec:apdx:hallucination_annotation}
We used the annotation workflow described in Section~\ref{ssec:hallucination_taxonomy} to guide the process of identifying and labeling hallucinations. Detailed definitions for each node (both non-hallucination and hallucination classes) are provided in Table~\ref{tab:annotation_label_definition}. 

To help annotators understand the essential elements of commit messages and code review comments, task definitions were also provided in ~\ref{ssec:apex:elements_in_code_review_commit_messages}. Through initial pilot rounds and discussions among annotators, we distilled a set of rules to guide the annotation process, which is provided in ~\ref{ssec:rules_annotation}.

\subsection{Essential Elements in Code Reviews and Commit Messages}
\label{ssec:apex:elements_in_code_review_commit_messages}

\paragraph{Code Review Comments} The primary purpose of code review comments is to offer constructive feedback from reviewers to code authors, aiming to improve code quality and maintain coding standards. A review comment often covers three elements: 

\begin{itemize}
    \item What (Evaluation): A review comment should point out what is the concern or issue in the code \cite{Yang-etal-2023-EvaCRC}.
    \item How (Suggestion): An ideal review comment provides suggestions for correction or prevention since code review is expected to help fix defects, improve quality, and address developers’ quality concerns \cite{Yang-etal-2023-EvaCRC}.
    \item Why: Explain the reasoning behind the concern and/or the suggested improvement \cite{Lin-etal-2024-Experience}.

\end{itemize}

\paragraph{Commit Messages} The primary purpose of commit messages is to provide developers (both current and future) with a summary of code changes, enabling them to understand how the code of a project has changed and why. Two elements have been shown to be essential for a commit message \cite{Liu-etal-2021-Neural,tian2022makes}.
\begin{itemize}
    \item What (Changes): A summary of what changes were made in the code. This often includes: 
    \begin{itemize}
        \item A summary of code object change that shows the object of change, characteristics of changes, or contrast before and after.  For example, \textit{``this commit removes the following deprecated properties: * ‘server.connection-timeout’ * ‘server.use-forward-headers’ [...]”}. Another example, \textit{``rename HeldCertificate.Builder.issuedBy() to signedBy()”}.
        \item An illustration of function. For example, \textit{Rename preferred-mapper property so its clear it only applies to JSON}) 
        \item Description of implementation principles. For example, \textit{“SslContextBuilder was using InetAddress.getByName(null) [...] On Android, null returns IPv6 loopback, which has the name ‘ip6-localhost’ ”}
    \end{itemize}
    \item Why: A justification of the motivation behind the code change. This often includes describing objectives or issues, illustrating requirements, or implying necessity.
    
\end{itemize}





\subsection{Summarized rules for annotation}
\label{ssec:rules_annotation}

\paragraph{Rules for Annotating Generated Code Reviews}
\begin{enumerate}
    \item  Unsure $\rightarrow$ Knowledge\_Overreach: a note of Knowledge\_Overreach should be left for cases that contain code snippets or software evolution (maintains, process related), we are not sure whether the generated content is true or not. E.g., \textit{``I think it would be better to use `getById` here.''}
    \item For a composite review that contains multiple sentences, there might be some sentences not functioning as review.  As long as there is at least one review exist, we consider it as review (not intent deviation). 
    \item   A review might have multiple sentences and each sentence has different labels, we decide the final label based on most severe one (label hallucination types if it exists). \\ For example, given this message \textit{``I think this is a bug. The `m\_indirectKernelMem` is a `std::vector<usm::memory>`. The `usm\_mem` is a single element of that vector. So this line is going to overwrite the `m\_indirectKernelMem` with a single element.”}. We have two labels: (a) we cannot tell that the \textit{m\_indirectKernelMem` is a `std::vector<usm::memory>} or not, which is `Unsure` requires fact checking; and (b) we know that \textit{``So this line is going to overwrite the `m\_indirectKernelMem` with a single element.''} is wrong based on the code context, it won't overwrite, so it's Input Inconsistency. Base on the two labels, we choose Input Inconsistency for this message. 

    \item How to distinguish it's a review or a justification? A review should contain the basic components of issue/concern, with optional suggestion and explanation, while a justification is a message aligned with the code change (no concern or suggestion, no new information inside).  For example, this message ``This is a bit of a hack, but I think it's the best we can do for now'' should be labeled as Intent Deviation since there is no any issue or concern. 
    \item Cases where the model suggests changing back to the older version without explanation, we don’t know whether the suggestion is better or not. If know exactly what to fact check, we label it Unsure (needs fact checking); otherwise, if it’s not violating the context, then we choose NO context deviation and then decide whether it's Informative or Uninformative. The following message should be labeled as Context Deviation $\rightarrow$ No and Informative, because it's sensible given the code context: ``I think this is a bit of a misnomer. I think it should be "Gets or sets JSON serialization settings".''. 
    \item In cases where the review is ambiguous, it might refer back to multiple places in the code patch, we label it as No-context deviation if it’s possible to apply in at least one kinds of scenario. Leave a comment of ``Can be interpreted as another wrong way''. In the example of: ``Layout/EmptyLinesAroundBlockBody: Extra empty line detected at block body end.'', where the `block body end' can be mapped to different places, one with an extra empty line and one without. 
    \item A review can apply to multiple places in the code patch, we prioritize mapping it to the code change part (-/+ lines) unless the review explicitly mentions other unchanged code snippets.  For example, in this message \textit{``I think this is a bit of overkill. We can just use `Fatal` and `Warning` directly.''}, the `Fatal' and `Warning' exist in both code changed parts and unchanged parts, but we prioritize the changed part. 
 
\end{enumerate}

\paragraph{Rules for Annotating Commit Messages}
\begin{enumerate}
    \item A message is considered as semantically equivalent to the ground truth message if the information you can get are equal after reading both. Specifically, both ``what'' changed in the code and and ``why'' it is changed should be aligned. 
    \item For semantic equivalence, we don’t not over-infer the meanings, if the message doesn’t explicit mention about it then it’s not. E.g., \textit{``Added support for CircleMarker''} we don't infer the CircleMarker is a type/instance of Marker unless the code explicitly defined it.
    \item For cases where we are not sure and cannot understand the message based on the given context, our prior knowledge and external web search, label it as Unsure, leave a note of “Difficult to comprehend the message”. 
    \item The <I> symbol comes from training data, where they mask out information referring to a different platform such as issue IDS, URLs, and version numbers. For example, the message \textit{``Bump to <I> (\#<I>)''} is not hallucinating, but it's  Uninformative based on the code change as it doesn't tell specifics of what bump to <I>. This message \textit{``removed unused imports from rfc<I>''} is considered informative based on the code context. 
\end{enumerate}


\subsection{red}{Examples with Annotated Hallucination Types}
\label{apdx:ssec:hallucination_examples}
We use the code review task as a representative example to illustrate the distinctions between different types of hallucinations.

\textbf{Input Inconsistency}
We annotate the model output code review \textit{``I think we should use \lstinline|Trim()| here as well.''} as an Input Inconsistency, since the corresponding code change (see below) already uses Trim(). While the comment may seem reasonable on its own, it conflicts with the actual code context and addresses a non-existent issue. 
\begin{lstlisting}[]
 @@ -144,6 +144,12 @@ namespace OpenTelemetry.Instrumentation.AspNetCore.Implementation 
     { 
         activity.SetTag(SemanticConventions.AttributeHttpUserAgent, userAgent); 
     } 
+ 
+                var xForwardedFor = request.Headers["X-Forwarded-For"].FirstOrDefault(); 
+                if (!string.IsNullOrEmpty(xForwardedFor)) 
+                { 
+                    activity.SetTag(SemanticConventions.AttributeHttpClientIP, xForwardedFor.Split(',').First().Trim()); 
+                } 
             } 
         } 
         
\end{lstlisting}

\textbf{Logic Inconsistency} We annotate the following model output as Logic Inconsistency: \textit{``I think it would be better to test that \lstinline|`resp.json['data']['data']['fingerprint']`| is a \lstinline|`dict`| and not a \lstinline|`dict`|.''}. The logic inside the review comment itself is problematic, regardless of the code input.

\textbf{Input Repetition}
For the code change below, the model output \textit{`input[form] polyfill, cannot load conditionally`} is annotated as \textit{Input Repetition}, as the output merely copies snippets from the code change.

\begin{lstlisting}[language=php]
@@ -15,6 +15,7 @@ return array(
         'vendor/bootstrap.min.js',
         'vendor/bootstrap-accessibility.min.js',
         'vendor/validator.min.js',
+        'vendor/form-attr-polyfill.js', // input[form] polyfill, cannot load conditionally, since we need all versions of IE
         'autocomplete.js',
         'common.js',
         'lightbox.js',
\end{lstlisting}
\textbf{Intent Deviation}

For the code change below, the model output \textit{``This is a bug fix. The \lstinline|`period.fraction`| was being used instead of \lstinline|`period.nanosec`| in the \lstinline|`DeadlineQosPolicy`| constructor.''} is annotated as \textit{Intent Deviation}, as it describes the code change rather than providing a review comment. It fails to identify issues or offer suggestions, thus violating the task objective.

\begin{lstlisting}[language=C++]
    @@ -42,7 +42,7 @@ bool DeadlineQosPolicy::addToCDRMessage(CDRMessage_t* msg)
     bool valid = CDRMessage::addUInt16(msg, this->Pid);
     valid &= CDRMessage::addUInt16(msg, this->length);//this->length);
     valid &= CDRMessage::addInt32(msg,period.seconds);
-    valid &= CDRMessage::addUInt32(msg,period.fraction);
+    valid &= CDRMessage::addUInt32(msg,period.nanosec);
     return valid;
 }
 
\end{lstlisting}

\section{Prompting and Fine-tuning Models}
\label{sec:apdx:prompting_fine_tuning_models}
\paragraph{Zero-shot prompting}

We use vLLM\footnote{\url{https://docs.vllm.ai/en/latest/}} for zero-shot prompting. The model temperature was set to 0 to make the output deterministic. We used the following prompts for code review and commit message generation.

\begin{tcolorbox}[colback=gray!10, colframe=gray!80, boxrule=0.5pt, arc=2pt, left=4pt, right=4pt, top=4pt, bottom=4pt]
Below is a code diff submitted during a code review process. \\
Please write a commit message within 50 words.

\texttt{[code\_diff]}: \texttt{\{code\_diff\}}

Respond only with valid JSON. Do not write an introduction or summary.
\end{tcolorbox}

\begin{tcolorbox}[colback=gray!10, colframe=gray!80, boxrule=0.5pt, arc=2pt, left=4pt, right=4pt, top=4pt, bottom=4pt]
    Below is a code diff submitted during a code review process. Please write a code review comment within 50 words to identify the concerns and suggest improvements.
    
    \texttt{[code\_diff]}: \texttt{\{code\_diff\}}
    
Respond only with valid JSON. Do not write an introduction or summary.
\end{tcolorbox}

\paragraph{Fine-tuning models}

We fine-tuned the three models on task-specific training data, including two general language models (Llama3.1-8B-Instruct\footnote{\url{https://huggingface.co/meta-llama/Llama-3.1-8B-Instruct}} and Qwen2.5-7B-Instruct\footnote{\url{https://huggingface.co/Qwen/Qwen2.5-7B-Instruct}}) and one specialized small language model pre-trained on code and commit message generation \cite{Lin2023CCT5}. The experiment was conducted on 1 NVIDIA H100 GPU. 

For CCT5~\cite{Lin2023CCT5}, we reused the code and original scripts from their replication package\footnote{\url{https://github.com/Ringbo/CCT5}} to fine-tune the model on our dataset. {The hyperparameters are: train\_batch\_size= 32, learning\_rate = 3e-4, max\_source\_length  = 512, max\_target\_length = 128 and warmup\_steps = 500, gradient\_accumulation\_steps = 4, maximum\_train\_steps = 150000, optimizer=AdamW.}



For LLaMA3.1-8B-Instruct and Qwen2.5-7B-Instruct, we perform instruction fine-tuning to further update the models parameters for the tasks at hand. We use full fine-tuning rather than parameter-efficient methods such as LoRA, as our preliminary experiments found that full fine-tuning performed better. The following instruction templates are used during training:

\begin{tcolorbox}[colback=gray!10, colframe=gray!80, boxrule=0.5pt, arc=2pt, left=4pt, right=4pt, top=4pt, bottom=4pt]
     Below is an instruction that describes a task, paired with an input that provides further context. Write an Output that appropriately completes the request.

    \#\#\# Instruction:
    Review the code diff and provide a constructive comment highlighting any issues and suggesting improvements.

    \#\#\# Input:
    
    Code diff: \texttt{\{code\_diff\}}

    \#\#\# Output:

    \texttt{\{code\_review\}}
    
\end{tcolorbox}

\begin{tcolorbox}[colback=gray!10, colframe=gray!80, boxrule=0.5pt, arc=2pt, left=4pt, right=4pt, top=4pt, bottom=4pt]
     Below is an instruction that describes a task, paired with an input that provides further context. Write an Output that appropriately completes the request.

    \#\#\# Instruction:
    You are a programmer who makes the below code changes. Please write a commit message for the below code diff
    
    \#\#\# Input:
    
    Code diff: \texttt{\{code\_diff\}}

    \#\#\# Output:
    
    \texttt{\{commit\_message\}}
    
\end{tcolorbox}
\begin{table*}[!t]
\centering
\scriptsize
\begin{tabular}{llcccc}
\hline
\textbf{Setting} & \textbf{Model} & \multicolumn{2}{c}{\textbf{CodeReview}} &  \multicolumn{2}{c}{\textbf{CommitBench}} \\
~ & ~ & Overall & Sample & Overall & Sample \\ 
\hline
\multirow{4}{*}{Zero-shot prompt} 
& Llama3.1-8B-Instruct       & 4.22 & 3.28 & 9.21 & 8.89 \\
& Qwen2.5-7B-Instruct   & 4.70 & 4.00 & 8.99 & 8.62  \\
& Llama3.1-70B-Instruct      & 3.88 & 4.09 & 9.72 & 9.88\\
& Qwen2.5-72B-Instruct  & 4.29 & 4.31 & 8.62 & 8.06 \\
\hline
\multirow{3}{*}{Fine-tuned}
& Llama3.1-8B-Instruct       & 5.28 & 5.25 & 15.06 & 15.29\\
& Qwen2.5-7B-Instruct   & 5.43 & 5.73 & 15.37 & 15.57 \\
    & CCT5                       & \textbf{5.58} & \textbf{6.53} & \textbf{17.45} & \textbf{17.46} \\
\bottomrule
\end{tabular}
\caption{Performance (BLEU-4 measured in \%) comparison of different models on CodeReview and CommitBench benchmarks under zero-shot and fine-tune settings. 
}
\label{tab:code_review_commitbench_bleu}
\end{table*}

{Regarding the hyperparameters used to fine-tune the two LLMs (Llama3.1-8B-Instruct and Qwen2.5-7B), we set the learning\_rate = 5e-5, max\_sequence\_length = 1024, batch\_size = 4. We set the max\_steps of fine-tuning to be 30000 and choose the best performing model on the validation set. The optimiser is Adamw.}

\paragraph{Results} We evaluated seven models in total, including four zero-shot and 2 fine-tuned models,\footnote{We consider the fine-tuned LLMs as different models from the ones before fine-tuning, as their weights have been updated for the tasks.} on their capability of generating task-specific messages using the traditional BLEU-4 metric \cite{papineni-etal-2002-bleu}. Table~\ref{tab:code_review_commitbench_bleu} presents the experimental results on code review comment generation and commit message generation across prompting and fine-tuning approaches.

The experimental results reveal several key patterns. First, zero-shot prompting approaches consistently underperform fine-tuned models, with BLEU scores ranging from 3.88-4.70\% for code review and 8.62-9.72\% for commit messages. In contrast, fine-tuned models achieve substantially higher performance, with the specialized CCT5 model reaching 5.58\% on code review and 17.45\% on commit messages. This highlights the necessity of fine-tuning for generating higher-quality \ccnl messages.

Second, code review proves to be a more challenging task compared to commit message generation, with BLEU scores approximately 2-3 times lower across all model configurations. This is sensible given that code review comments require models to critically analyze and provide constructive feedback on code changes, representing a higher cognitive demand than the descriptive nature of commit messages.

The performance on our manually sampled subset closely mirrors the overall dataset performance, with sample BLEU scores showing similar trends (e.g., CCT5 achieving 6.53\% vs 5.58\% overall for code review), validating the representativeness of our evaluation approach.

\section{Examples of Code Changes}
\label{apdx:patch_examples}

\subsection{Example from CodeReview}
\label{apdx:patch_examples_codereview}
\begin{tcolorbox}[colback=gray!10, colframe=gray!80, boxrule=0.5pt, arc=2pt, left=4pt, right=4pt, top=4pt, bottom=4pt]
\begin{lstlisting}[language=Java]
@@ -65,6 +65,7 @@ public class SmartStorePlugin extends ForcePlugin {
 	public static final String LIKE_KEY = "likeKey";
 	public static final String MATCH_KEY = "matchKey";
 	public static final String SMART_SQL = "smartSql";
+	public static final String ORDER_PATH = "orderPath";
 	public static final String ORDER = "order";
 	public static final String PAGE_SIZE = "pageSize";
 	public static final String QUERY_TYPE = "queryType";
\end{lstlisting}

\#\#\# Human Code Review:
Previously you could only order by the field you were searching by. But for full-text search, you can search across all indexed fields, and it didn't make sense not to have a sorting. For backward compatibility, the javascript code uses indexPath as the orderPath when no orderPath is provided.

\#\#\# CCT5 Code Review: 
I think this should be \texttt{orderPath} instead of \texttt{orderPathKey}.
\end{tcolorbox}

\subsection{Example from CommitBench}
\label{apdx:patch_examples_commitbench}

\begin{tcolorbox}[colback=gray!10, colframe=gray!80, boxrule=0.5pt, arc=2pt, left=4pt, right=4pt, top=4pt, bottom=4pt]
\begin{lstlisting}[language=Go]
diff --git a/nomad/server.go b/nomad/server.go
index <HASH>..<HASH> 100644
--- a/nomad/server.go
+++ b/nomad/server.go
@@ -1169,7 +1169,12 @@ func (s *Server) setupRaft() error {
 			}
 		} else if _, err := os.Stat(peersFile); err == nil {
 			s.logger.Info("found peers.json file, recovering Raft configuration...")
-			configuration, err := raft.ReadPeersJSON(peersFile)
+			var configuration raft.Configuration
+			if s.config.RaftConfig.ProtocolVersion < 3 {
+				configuration, err = raft.ReadPeersJSON(peersFile)
+			} else {
+				configuration, err = raft.ReadConfigJSON(peersFile)
+			}
 			if err != nil {
 				return fmt.Errorf("recovery failed to parse peers.json: %v", err)
 			}
\end{lstlisting}

\#\#\# Human Commit Message: Add support in nomad for supporting raft 3 protocol peers.json

\#\#\# CCT5 Commit Message: 
nomad: fix peers.json recovery for protocol version 3
\end{tcolorbox}

\section{Hallucination Detection}
\subsection{Hallucination Detection Methodology Details}
\label{ssec:apdx:hallu_detection_methods}
We adopt existing hallucination measurement metrics, including reference-based and reference-free hallucination detection approaches to address different practical needs. Reference-based metrics serve as valuable benchmarks during model training and evaluation when gold standards are available, while reference-free methods enable hallucination detection in real-world deployment scenarios where reference texts are typically unavailable. 


\subsubsection{Reference-based Metrics}
In reference-based metrics, hallucination is estimated by the quality of a generation $y$, which is evaluated by comparing against the reference $\hat{y}$ using certain metrics. The hypothesis is that the lower the quality is, the more likely $y$ it is to be a hallucination. We use two metrics that are widely used for quality estimation: Lexical overlap with BLEU, and Natural Language Inference. 

\textbf{Lexical overlap} metrics such as {BLEU} evaluate the n-gram overlap between the $y$ and $\hat{y}$. This type of metric has been widely used in prior work to evaluate the quality of generated commit messages \cite{liu2018neural,li2024only} and review comments \cite{tufano2021towards,Li2022CodeReviewer}. Recently, it has also been adapted to study the correlation with hallucinations in natural language generation tasks, such as machine translation \cite{guerreiro-etal-2023-looking,dale-etal-2023-detecting}. 

\textbf{Natural Language Inference (NLI).} NLI is a standard NLP task that evaluates the logic relationship between a pair of premise and hypothesis sentences, determining whether it is entailment, contradiction, or neutral, which has been widely used to evaluate the factual consistency \cite{hu2024refchecker, valentin2024cost} and hallucination detection \cite{manakul-etal-2023-selfcheckgpt, elaraby2023halo}. We use NLI to measure the probability of the reference $y$ entails the the generated NL $\hat{y}$. The intuition is that if the $y$ can be directly inferred from the reference $\hat{y}$, then it is high quality and less likely to hallucinate. 
We used the best performing model nli-deberta-v3\footnote{https://huggingface.co/cross-encoder/nli-deberta-v3-base} based on the performance on Sentence Transformer \footnote{https://sbert.net/} to obtain the entailment logit. 





\subsubsection{Reference-free Metrics}
In reference-free measurements, reference is not accessed, only information  from the source input or from the model behaviors while generating a sequence is used. We use three types of measurements: similarity-based, uncertainty-based, and feature-attribution based. 


\paragraph{Similarity between the generation and the source}
We estimate semantic similarity between source and generation using cosine similarity $\cos(E_y, E_x)$ between embeddings of generated NL $y$ and source code $x$. The intuition is that irrelevant generations are less similar and more likely to hallucinate.  To obtain the embeddings, we use three models pre-trained on both code and natural language corpora: codebert-base\footnote{https://huggingface.co/microsoft/codebert-base}, codet5p-220m-bimodal\footnote{https://huggingface.co/Salesforce/codet5p-220m-bimodal}, and codet5p-770m\footnote{https://huggingface.co/Salesforce/codet5p-770m}.

\paragraph{Sequence-level confidence scores}
A sequence-level confidence score has been used in machine translation for hallucination detection \cite{guerreiro-etal-2023-looking,huang2023look}, where it is calculated via aggregating token-level uncertainty into sentence level by taking the average across the sequence. Token-level confidence can be measured in various ways. The intuition is when a model hallucinates, it tends to be less confident. Several metrics have been proposed to estimate the token-level uncertainty, including probability, logit and entropy \cite{guerreiro-etal-2023-looking, huang2023look, valentin2024cost}. 

We also use entropy to measure uncertainty: a more uniform token distribution (higher entropy) indicates lower model certainty. This can be formulated as follows:
\begin{align}
    \label{eq:seqeng}
    \text{SeqEntropy} = \frac{1}{L}\sum_{i=1}^{L} {H_i},
\end{align}
where $H_i$ is the entropy of the token distribution.

\paragraph{Feature attribution}  
In a transformer-based model $M$, generating a token $y_t$ involves both the input $x$ and previously generated target tokens ($y_1$ to $y_{t-1}$). Prior work has shown that the interaction between $y_t$ and these sources reveals hallucination patterns \cite{tang2022reducing,chen2025attributive,Snyderetal2024-Early}, which can be detected through feature attribution in NL hallucinations.
We conduct both feature attribution for both the input source $x$ and the previously generated target tokens. 


We employ a widely used feature attribution method Input X Gradient~\cite{pmlr-v70-shrikumar17a}, which calculates the gradient of the output with respect to the input and considers the impact of input magnitudes on generation. The attribution score from $x_i$ to $y_t$ can be formulated as:
\begin{align}
A_{i,t} = x_i \times \frac{\partial y_t}{\partial x_i}
\end{align}
where $A_{i,t}$ is the attribution score, and $\frac{\partial y_t}{\partial x_i}$ denotes the gradient of $y_t$ in an attribution model $M$ with respect to the input $x_i$. A higher $A_{i,t}$ indicates that $x_i$ is more important for generating $y_t$.



\textbf{Source Attribution Score.} To investigate hallucinations on sequence level, we apply an aggregation function on $A$ to convert a sequence of token-level attribution scores into a single attribution value. We first compute the maximum attribution value across all input tokens for each output token $y_t$, then take the average of these maximum values. The attribution score of the source to the generated sequence. 
\begin{align}
    \text{SourceAttr} = \frac{1}{T}\sum_{t=1}^{T} \max_{i \in [1,N]} A_{i,t},
\end{align}

where $T$ is the length of the generated sequence, $\text{SourceAttr}$ represents final sequence-level overall source contribution score. The intuition is that when the maximum input contribution is small, the generated $y$ is likely to be a hallucination as the model didn't generate based on the input. 

Given our input is a code change consisting of both old and new code, human developers primarily focus on the changed parts when generating commit messages and code review comments. Based on this observation and the assumption that models should similarly emphasize code changes, we designed variations of the aggregation methods that separate attribution scores for changed and unchanged code. Our hypothesis is that lower attribution scores on the changed parts indicate a higher likelihood of hallucination.
\begin{align}
    \text{ChangedAttr} &= \frac{1}{T}\sum_{t=1}^{T} \max_{i \in C} A_{i,t}, \\
    \text{UnchangedAttr} &= \frac{1}{T}\sum_{t=1}^{T} \max_{i \in [1,N] \setminus C} A_{i,t}, 
\end{align}
where $C \subset [1,N]$ represents the indices of tokens in the changed code (all - and + lines), and $[1,N] \setminus C$ represents the indices of unchanged code tokens.

\textbf{Target Attribution.} We also calculate the attribution score from previously generated tokens:
\begin{align}
\text{TargetAttr} &= \frac{1}{T}\sum_{t=1}^{T} \max_{j \in {1, \dots, t-1}} \hat{A}{j,t},
\end{align}
where $\hat{A}{j,t}$ is the attribution score from $y_1$ to $y_j$ ($j$ ranges from 1 to $t-1$). The final TargetAttr score denotes the overall maximum attribution score from previously generated tokens to the current token.





To obtain attribution scores for generated sequences, we use constrained attribution~\cite{sarti-etal-2023-inseq} through the Inseq library.\footnote{https://inseq.org/en/latest/} 
Constrained attribution works by providing an attribution model $M$ with both the input code $x$ and the generated output $y$, then analyzing how the model associates each input token with each output token step by step. Rather than generating text freely, the model is constrained to follow the specified target sequence, allowing us to measure which parts of the input most strongly influence each token in the output. This reveals the model’s implicit justification for each output token based on the input.

As the attribution model $M$, we use the same three models fine-tuned in our RQ1 experiments for each task: LLaMA3.1-8B-Instruct, Qwen2.5-7B-Instruct, and CCT5. For each generation, we apply both self-attribution (where the generator attributes its own output, e.g., CCT5 attributes its own generation) and cross-attribution (where a different model attributes the output, e.g., CCT5 attributes LLaMA3.1-8B's generation). This dual perspective helps us understand whether a model is aware of its own hallucinations and whether external models can detect hallucinations based on attribution signals. While attributing each output token, we also extract uncertainty scores based on logit, probability, and entropy. 

\subsection{Complementarity Among Individual Detection Metrics}
\label{apdx:sec:metric_complementarity}

To examine how different types of metrics complement each other, we select the top three individual metrics (one from each category) based on ROC-AUC.

For {CodeReviewer}, we choose {logit\_Llama3.1}, {similarity\_score\_codebert-base}, and {changed\_contribution\_CCT5}.  
For {CommitBench}, we select {similarity\_score\_codet5p-770m}, {target\_target\_contrib\_CCT5}, and {logit\_Llama3.1}.

From each metric, we extract the top 25\% samples ranked by their metric score, indicating that they are highly correlated with hallucination labels. We then analyze the overlaps and unions of these sets.

Figure~\ref{fig:venn_diagram_three_metrics} shows the Venn diagrams of the selected metrics. On \textsc{CodeReviewer}, the three metrics capture almost disjoint sets. On \textsc{CommitBench}, only three samples are shared across all three metrics, suggesting strong complementarity.



\subsection{Correlation between Detection Metrics and Hallucination}
\label{ssec:apdx:point_biserial_correlation}

In addition to ROC-AUC, we also analyzed the correlation between each individual metric and the hallucination labels we annotated (hallucination = 1, non-hallucination = 0). To evaluate the correlation, we use the point-biserial correlation coefficient ($r_{pb}$), which measures the strength and direction of the relationship between a continuous variable (i.e., metric scores) and a dichotomous variable (i.e., the binary hallucination label).

\begin{figure*}[!t]
    \centering
    \includegraphics[width=\linewidth]{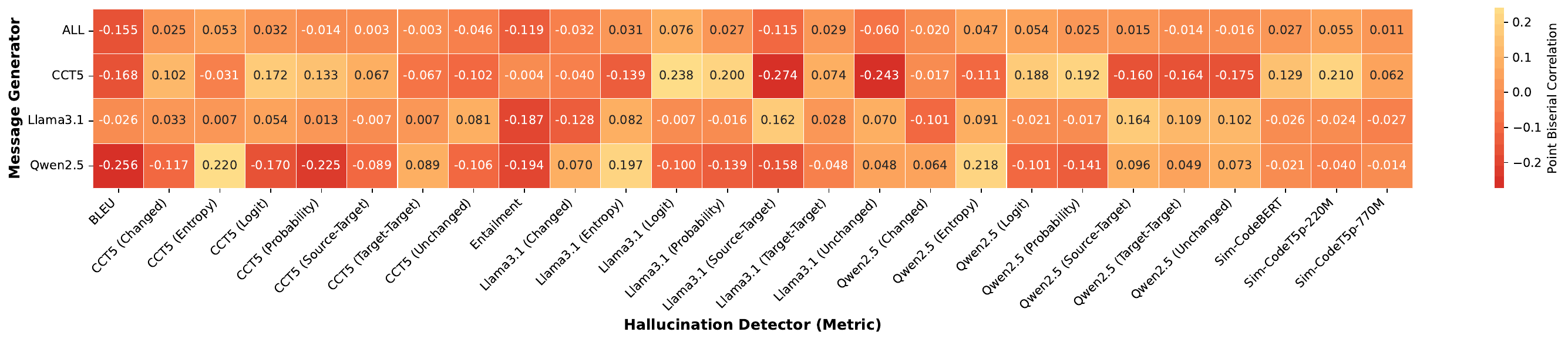}
    \caption{Point-biserial correlation between metrics and hallucinations on CodeReviewer.}
    \label{fig:hallu_point_biserial_corr_coderviewer}
\end{figure*}

\begin{figure*}[!t]
    \centering
    \includegraphics[width=\linewidth]{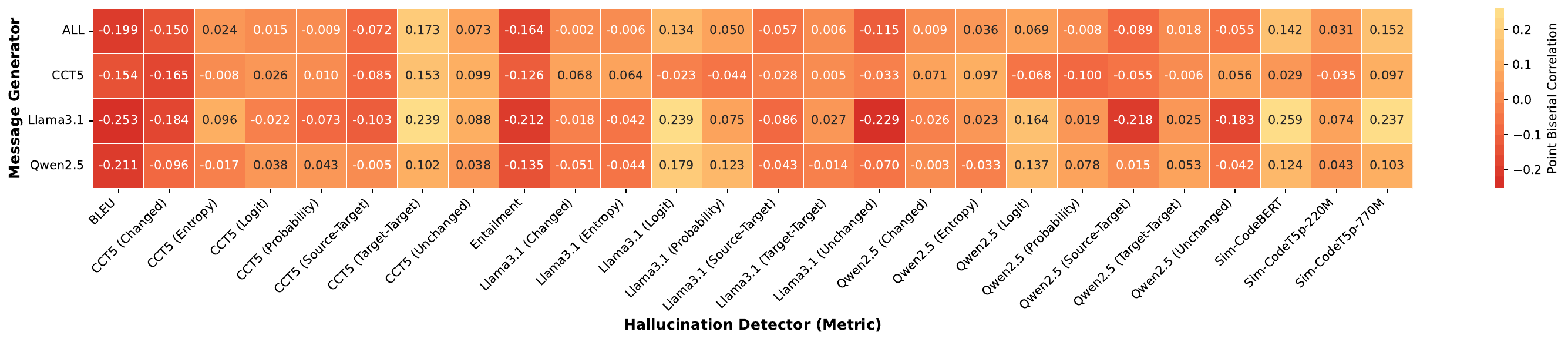}
    \caption{Point-biserial correlation between metrics and hallucinations on CommitBench.}
    \label{fig:hallu_point_biserial_corr_commitbench}
\end{figure*}

The results are presented in Figures~\ref{fig:hallu_point_biserial_corr_coderviewer} and~\ref{fig:hallu_point_biserial_corr_commitbench}. Overall, the correlation is weak ($|r_{pb}| \in [0, 0.2)$) across all samples for individual metrics. However, when examining generator-specific results, the correlation between certain generator–metric pairs increases ($|r_{pb}| \in [0.2, 0.3)$).

These findings further motivate our exploration of how combining multiple metrics can improve hallucination detection.



\begin{figure*}
    \centering
    \includegraphics[width=\linewidth]{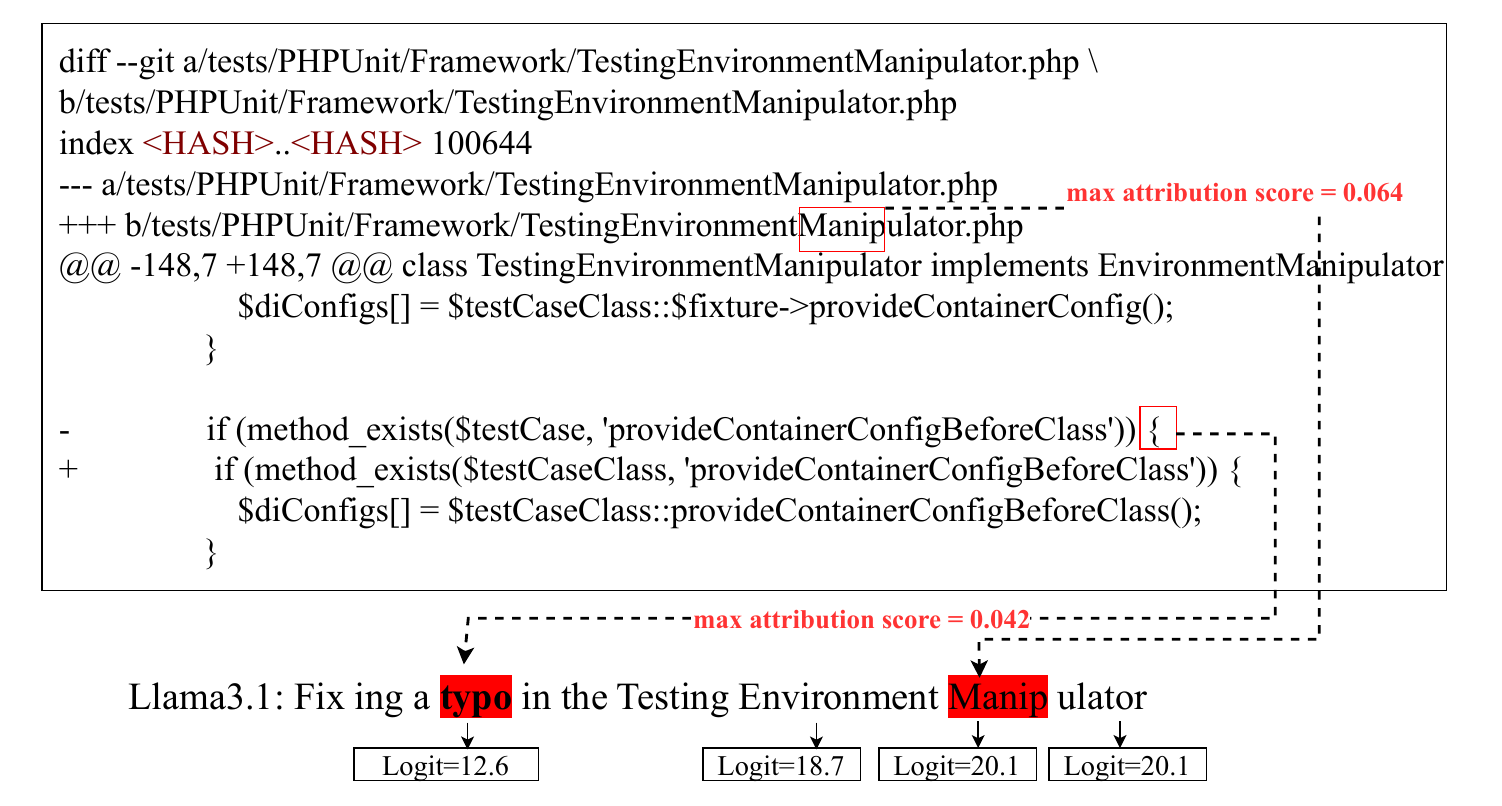}
    \caption{An example of feature attribution on a hallucinated commit message comment generated by Llama3.1. Attribution model: Llama3.1.}
    \label{fig:comit_bench_case_study}
\end{figure*}

\subsection{Signs of Coefficients in LR model}
\label{apdx:ssec:lr_opposite_sign_coef}

\begin{figure}[!ht]
    \centering
    \includegraphics[width=\linewidth]{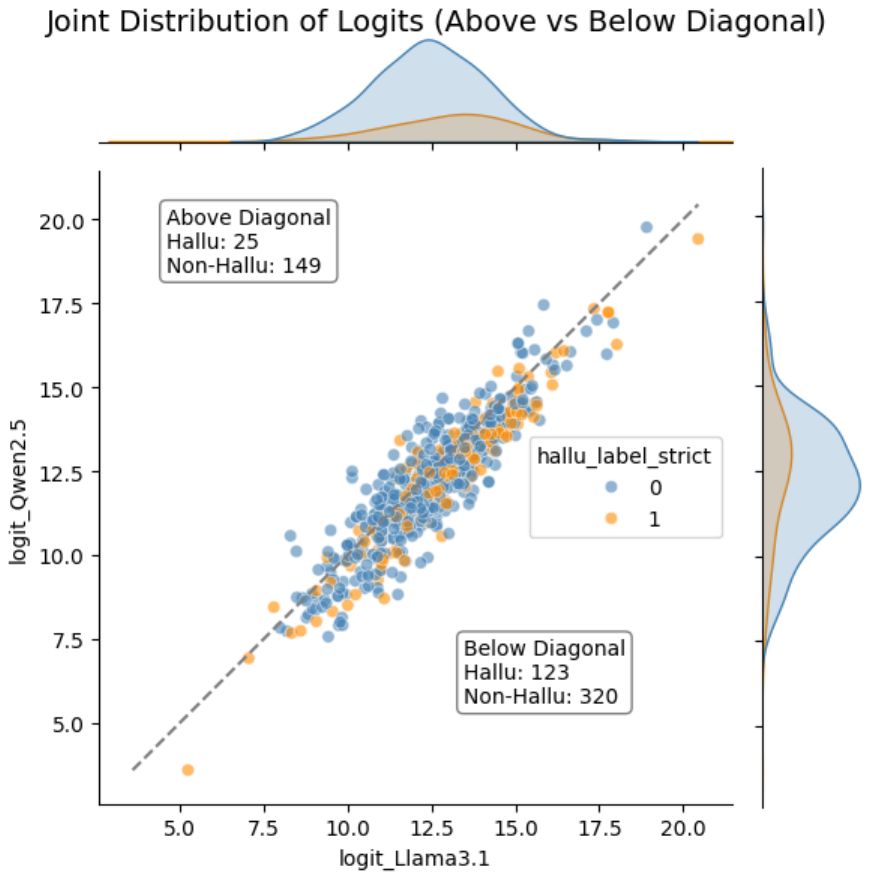}
    \caption{Joint Distribution of Qwen and Llama Logits on CommitBench dataset.}
    \label{fig:qwen_llama_logit_sign_commitbench}
\end{figure}

In Section~\ref{ssec:combien_multiple_metrics} (Table~\ref{tab:rq2_logist_regression_commitbench}), we observed that the two uncertainty-based metrics—{logit\_Llama3.1} and {logit\_Qwen2.5}—both contribute significantly to hallucination prediction, but with opposite coefficient signs: positive for {logit\_Llama3.1} and negative for {logit\_Qwen2.5}.
The signs of the coefficients indicate that higher logits from LLaMA3.1 are associated with hallucinations, whereas higher logits from Qwen2.5 are associated with non-hallucinations.
We hypothesize that Qwen’s confidence is more reliable, while LLaMA3.1 tends to be overconfident.
To further explore this, we plot the joint distribution of the two logits in Figure~\ref{fig:qwen_llama_logit_sign_commitbench}.
When Qwen2.5 is more confident than LLaMA3.1 (above the diagonal), hallucinations are less frequent; conversely, when LLaMA3.1 is more confident (below the diagonal), hallucinations occur more often.
This pattern supports our hypothesis.

This observation aligns with prior work~\cite{zhou-etal-2023-navigating, mielke-etal-2022-reducing}, which shows that models can be overconfident when generating outputs due to differences in training data and strategies. In our study, both models were fine-tuned on the same data, so we suspect this difference is partly due to pre-training.


\subsection{LR Model Predictions by Hallucination Type}

To understand which hallucination types are correctly detected, we examine samples predicted as hallucinations by our best logistic regression models on {CodeReviewer} and {CommitBench} (Section~\ref{sec:rq2}).

\begin{figure}[!t]
  \centering
  \begin{subfigure}[t]{0.44\textwidth}
    \centering
    \includegraphics[width=\linewidth]{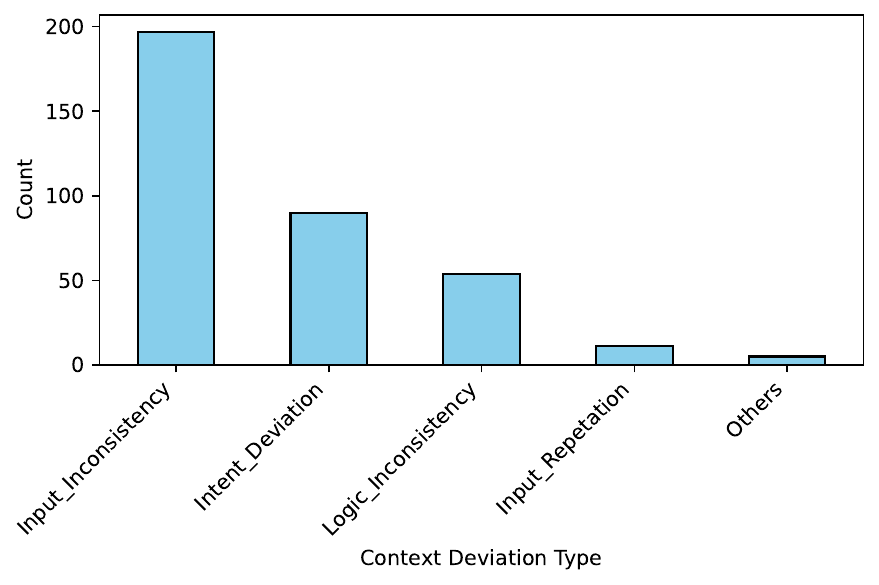}
    \caption{Codereviewer}
  \end{subfigure}
  \vspace{0.5pt}
  \begin{subfigure}[t]{0.44\textwidth}
    \centering
    \includegraphics[width=\linewidth]{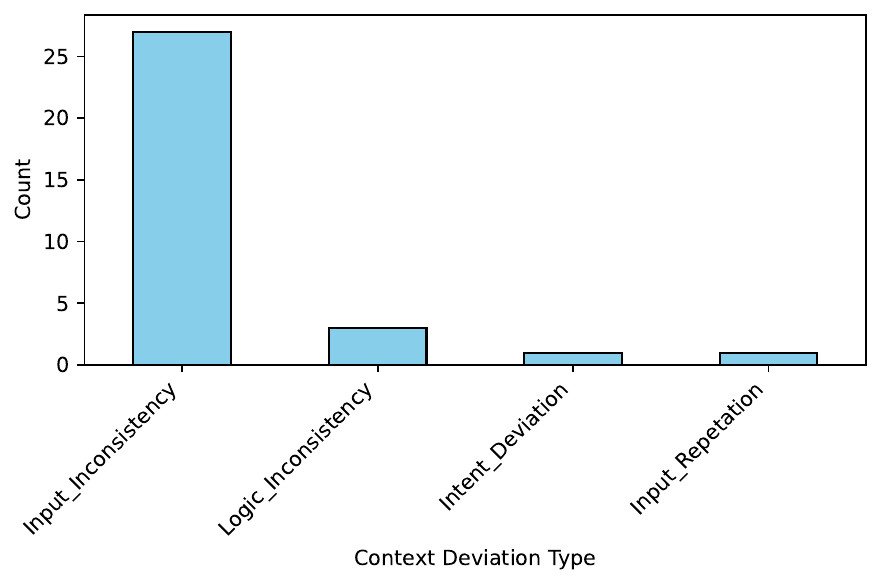}
    \caption{Commitbench}
  \end{subfigure}
  \caption{hallucination type distribution on LR models corrected predicted as hallucination.}
  \label{fig:bar_distribution_lr_pred_hallu_type}
\end{figure}

Figure~\ref{fig:bar_distribution_lr_pred_hallu_type} shows the type distributions. They largely mirror the overall dataset distribution, with \textsc{Input Inconsistency} most frequent in both datasets, followed by \textsc{Intent Deviation} in \textsc{CodeReviewer}, and \textsc{Logic Inconsistency} thereafter.

\subsection{{LR model prediction per programming language}}
\label{apdx:ssec:per_programming_language}
While our hallucination detection approach is language-agnostic, model performance may still be influenced by programming language distributions in pre-training and fine-tuning data. To examine this, we analyze the distribution of programming languages among samples predicted as hallucinations by the logistic regression model and compare it to the distribution of samples labeled as hallucination in the full test set.

\begin{figure}[!t]
  \centering
  \begin{subfigure}[t]{0.44\textwidth}
    \centering
    \includegraphics[width=\linewidth]{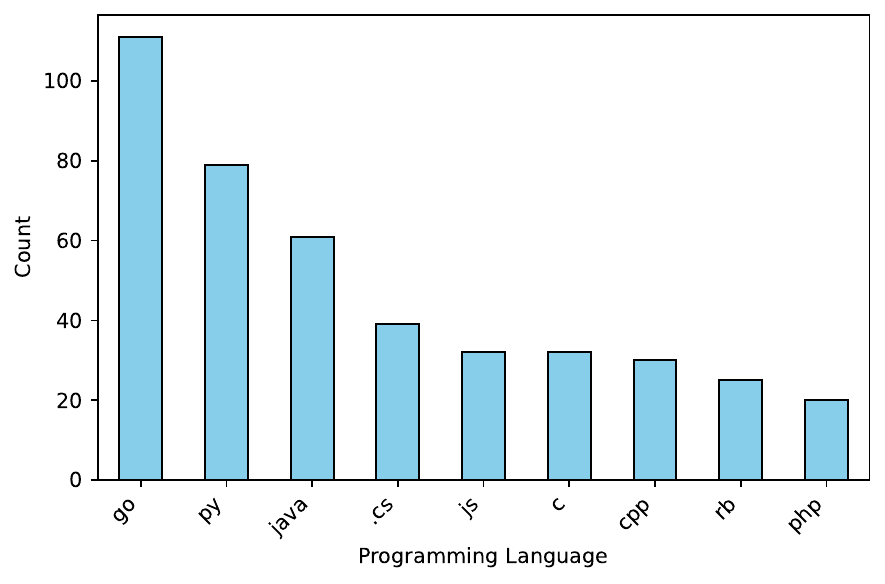}
    \caption{Samples in model corrected predicted as hallucination }
  \end{subfigure}
  \vspace{0.5pt}
  \begin{subfigure}[t]{0.44\textwidth}
    \centering
    \includegraphics[width=\linewidth]{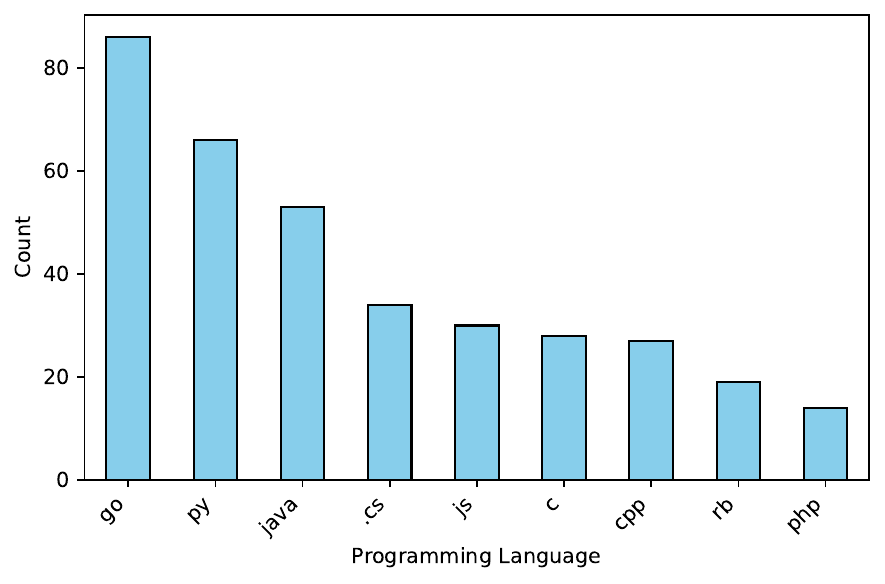}
    \caption{Samples in test set}
  \end{subfigure}
  \caption{CodeReviewer: programming language distribution on the model corrected predicted as hallucinations and our test set.}
  \label{fig:bar_distribution_programming_language_codereviewer}
\end{figure}

\begin{figure}[!t]
  \centering
  \begin{subfigure}[t]{0.44\textwidth}
    \centering
    \includegraphics[width=\linewidth]{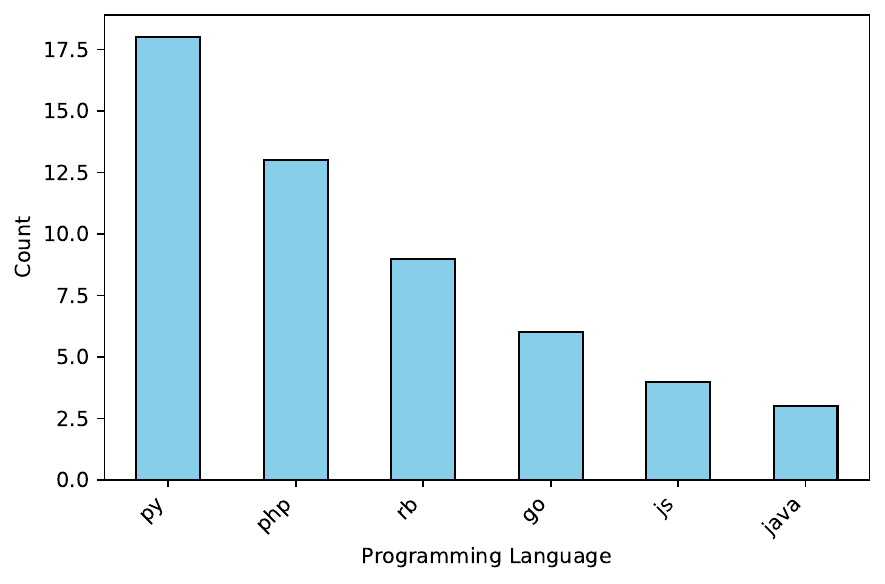}
    \caption{Samples in model corrected predicted as hallucination}
  \end{subfigure}
  \vspace{0.5pt}
  \begin{subfigure}[t]{0.44\textwidth}
    \centering
    \includegraphics[width=\linewidth]{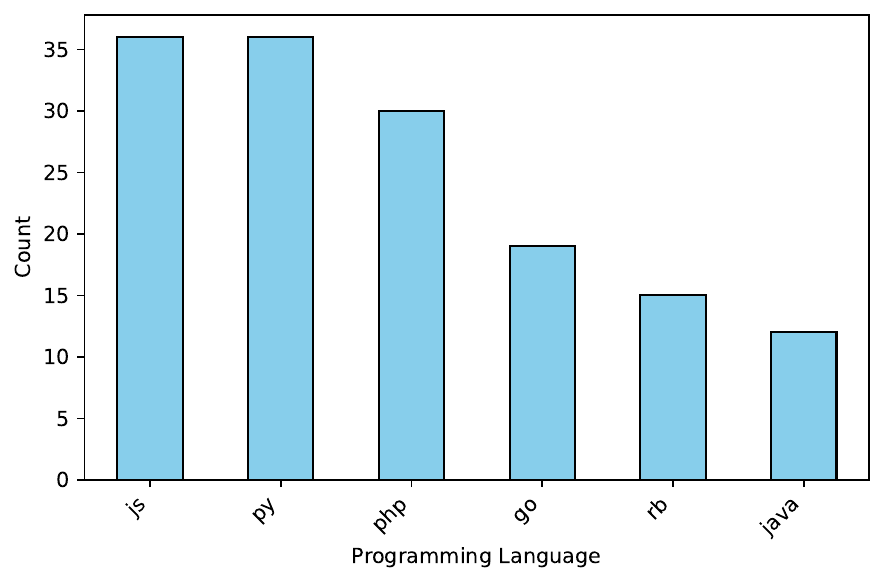}
    \caption{Samples in Test set}
  \end{subfigure}
  \caption{CommitBench: programming language distribution on the model corrected predicted as hallucinations and our test set.}
  \label{fig:bar_distribution_programming_language_commitbench}
\end{figure}

The results are shown in Figure~\ref{fig:bar_distribution_programming_language_codereviewer} for \textsc{CodeReviewer} and Figure~\ref{fig:bar_distribution_programming_language_commitbench} for \textsc{CommitBench}. In \textsc{CodeReviewer}, the language distribution of model predictions closely matches that of the test set, suggesting consistent detection across languages. In \textsc{CommitBench}, the distributions also largely align, with one notable exception: JavaScript (\texttt{js}) is the most dominant in the test set but is not predicted (recalled) in the model’s predicted hallucinations.


\end{document}